On the security and privacy of Interac e-Transfers

Fabian Willems
Mohammad Raahemi
MSc students in Electronic Business Technologies (EBT),
University of Ottawa
{fabian.willems, mohammad.raahemi}@uottawa.ca

Prasadith Buddhitha, MSc,
Ph.D. student in Electronic Business,
University of Ottawa

Carlisle Adams, Ph.D., P.Eng.,
Professor at School of Electrical Engineering and Computer Science (EECS),
University of Ottawa

Thomas Tran, Ph.D.,
Professor at School of Electrical Engineering and Computer Science (EECS),
University of Ottawa

Updated version as of July 15th, 2019
Reformatted for publication on December 10th, 2019

-Extended version-

School of Electrical Engineering and Computer Science
Faculty of Engineering
University of Ottawa

## Abstract

Nowadays, the Interac e-Transfer is one of the most important remote payment methods for Canadian consumers. To the best of our knowledge, this paper is the very first to examine the privacy and security of Interac e-Transfers. Experimental results show that the notifications sent to customers via email and SMS contain sensitive private information that can potentially be observed by third parties. Anyone with illegitimate intent can use this information to carry out attacks, including the fraudulent redirection of *Standard e-Transfers*. Such an attack is shown to be possible at least in an experimental setup but likely also in reality. Recent news articles support this finding. Improvements to overcome these interconnected privacy and security problems are proposed and discussed.

## Keywords

Interac e-Transfer, Email Money Transfer, Security, Privacy, Fraud, e-Transfer Redirection, Online Payment, Remote Payment, Canada

## 1. Introduction

Since its introduction in 2002, the steady growth of *Interac e-Transfer* volume has made it the third most frequently used remote payment method for Canadian consumers, finally overtaking its biggest rival in the field, paper & cheques [1]–[3]. But not only consumers use e-Transfers. Including business transactions, more than 371 million Interac e-Transfers were performed in 2018, a total value of more than CAD 132 billion. Compared to 2017, these numbers show an increase of 54 % in volume and 45 % in value [2], [4]. Moreover, 57 % of Canadians and actually 4 out of 5 Canadian online banking users are registered for the service at one of the 257 currently participating financial institutions [4], [5].

Clearly, such an important financial transfer system has to maintain the security and privacy of its transactions in order to mitigate the risk of fraud and disclosure of its users' private financial information. The Interac Corporation claims that "Interac e-Transfer users are protected with multiple layers of security, making the service one of the most secure money transfer services globally" [6]. These layers are detailed to be "Authentication and transaction encryption[,] Financial institution authentication [and] Proprietary risk management" [6]. The latter apparently incorporates a "World-class fraud scoring engine to ensure validity of sender, requestor and recipient" [7]. Additionally, the system allegedly "requires no bank account information, and no personal financial information of any kind, to move money between people or businesses" [8].

Regarding these claims, the authors of this paper were surprised to notice in their everyday banking that recipients and senders of Interac e-Transfers are notified and given status updates by emails containing information about the senders' and recipients' names, financial institutions and transferred amounts. To our even greater bewilderment, recipients of Interac e-Transfers are authenticated by often user-defined security questions. Hereby some senders choose easily answerable questions as "Where do I live?" or worse "What is my name?", which obviously should not be possible in a *secure money transfer service* or, more generally, an online banking environment. Based on this we conducted a first ad-hoc experiment successfully depositing CAD 0.10 in an account whose owner's legal name was not related to the specified addressee.

The above observations finally led to the following research questions.

Question 1:  Under which circumstances can Interac e-transfers be hijacked (redirected)?

Question 2:  In which way are Interac e-Transfer security questions insecure?

Question 3:  Do the recently introduced Autodeposit and Request Money features increase security and/or do they create new issues?

Question 4:  What can potential adversaries learn from email or SMS notifications?

Question 5:  How can the learned information be misused for more sophisticated attacks?

Question 6:  What are the reasons for the potential security and privacy problems we found, i.e. why has the system been designed as it is today?

To answer these questions, we have designed and performed two experiments in the form of two series of Interac e-Transfers. The first series focused on researching privacy related problems, the second on security related problems. However, as this paper will show, there is a close link between the identified security and privacy problems.

To the best of our knowledge, there was no – at least not publicly available – prior research of potential security and privacy problems regarding the Interac e-Transfer platform. Therefore, this paper firstly qualifies and quantifies these problems based on the analysis of our experimental results and publicly available information about the system. Because of research restraints and time limits (see section 3.1), our results – especially regarding the security problems – have to be seen as preliminary conclusions (see section 5).

Secondly, this paper proposes improvements to the Interac e-Transfer platform to address the identified problems. Accordingly, the remainder of this paper is organized as follows:

The next section (2) will give further necessary background information regarding the Interac e-Transfer platform and will briefly consider related work. Section 3 continues with a description of the performed experiments and a detailed analysis of the results regarding security and privacy of the system. In addition, further observations about the system are described and explained before possible improvements are discussed. An evaluation of the results and proposed improvements follow in Section 4. We discuss contributions and conclude in Section 5, also outlining future research. Finally, Section 6 gives information about our discussion with Interac regarding the results of this research.

A short version of this paper can be found at https://arxiv.org/abs/1910.01587.

## 2. Background and Related Work

### 2.1. History of Interac e-Transfer Companies and Brands

This paper will refer to the service as *Interac e-Transfer* or *e-Transfer* for short although other company names and brands were used in the past (for details, see Appendix A).

### 2.2. Types of Interac e-Transfers

The Interac e-Transfer platform differentiates between three types of e-Transfers for consumers [5]:

1. The *original Interac e-Transfer* (*Standard e-Transfer*) requires a manual selection of the financial institution and account into which transferred funds shall be deposited. The sender specifies a security question which the recipient must correctly answer to deposit the funds (verification of legitimacy).

2. The *Interac e-Transfer Autodeposit* (*Autodeposit e-Transfer*) is similar to the *Standard e-Transfer* but the recipient address is associated with a single bank account. Thus, the funds are automatically deposited, and a security question is not required.

3. An *Interac e-Transfer Request Money* (*Request Money e-Transfer*) can be regarded as requesting the initiation of a one-time ad-hoc Autodeposit: the recipient authorizes the request the sender sent which directly deposits the requested amount in the account specified by the sender. A security question is not required.

A detailed description of the different types of Interac e-Transfers can be found in Appendix B.

## 2.3. The Interac e-Transfer Platform Architecture

Figure 1 presents a high-level view of the Interac e-Transfer platform architecture, which is described in the following paragraphs.

As the previous sections show, Interac e-Transfers allow for transferring funds between customer accounts at Canadian financial institutions. The customers hereby mainly interact with their own financial institution via online or mobile banking. The Interac Corp. acts as a Central Clearing Facility which transfers the funds via sending and receiving them to and from trust accounts (suspense accounts) at the respective financial institutions [7, Sec. How It Works]. This means that the funds are instantly deducted from the sender's account when a transfer is started and that they are also instantly available in the recipient's account when the transfer is successfully completed. This approach, called *Good funds model,* therefore transfers funds with only slight delays (up to 30 minutes) within what is called "a private, real-time clearing network" [9, p. 7]. This is to be seen in contrast to other payment methods as e.g.

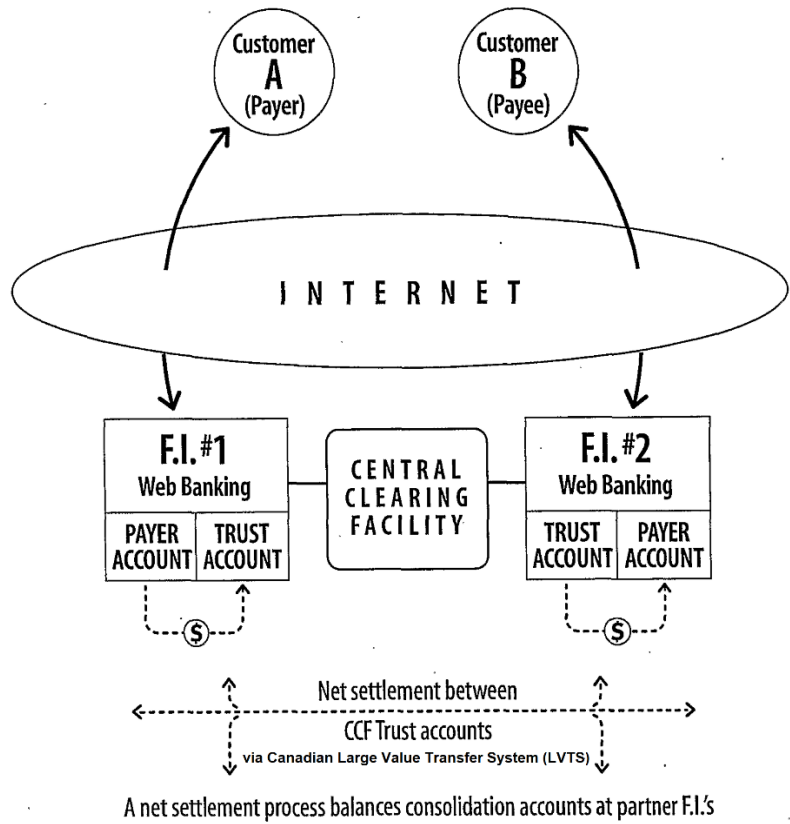

*Figure 1: Interac e-Transfer platform architecture, supplemented version, based on original taken from [9, Fig. 4a]*

cheques, which are based on the *Promise to pay model* and are cleared via slower clearing mechanisms so that transferred funds are withheld a much longer period of time [8, p. 3].[8], [9, p. 7]. The daily settling of fund differences in the trust accounts is nowadays technically performed via the Canadian *Large Value Transfer System (LVTS)* [7, Sec. How It Works], [9, p. 9].

From the perspective of the participating financial institutions (*participants*), the Interac Corp. not only acts as a Central Clearing Facility but also as an *Application Service Provider (ASP)* for Interac e-Transfers: the required functionalities are integrated into the participants' online and mobile banking services via an XML-based interface. The Interac Corp. therefore not only keeps track of the transferred funds but also of each e-Transfer's status. This includes the sending of notification emails or SMS text messages to the participants' customers, which, where necessary, contain a customized link to the customer's preferred financial institution [7, Sec. How It Works], [9]. In the case that the customer's preferred financial institution is not known and therefore not contained in the notification and to allow for selecting another financial institution, there is an additional custom link to the Interac e-Transfer web site. Here, the customer can choose from all participating financial institutions and is then taken to the respective financial institution's login page. Figure 2 gives an overview of the different interfaces being used in the Interac e-Transfer platform architecture.

According to Interac, the Interac e-Transfer platform has layers of security and fraud protection in place, but due to these being their confidential intellectual property, the authors of this paper have no insight into this and so they are neither described nor shown here [6], [7], [10]. However, this paper does not consider the backend of the platform but focuses on other parts as described in section 3.1. Therefore, we only consider whether these layers of security are sufficient to maintain security with respect to the experiments we performed. Additionally, the authors had to rely solely on publicly available information about the system. With regard to this, Interac claims that the patent describing the internal workings of the system is "outdated" [10]. In this regard, we believe it is likely that the system has been revised since its introduction and that, for example, (new) layers of security have been added. However, we also believe that the main architecture described here has not changed. This view seems to be confirmed by recent descriptions of the system [7], [8].

## 2.4. Related Work

Although Interac e-Transfers are an established and increasingly important payment method in Canada, there has been little to no – published – scientific research about its security and privacy so far. Databases of scientific publications such as Web of Science (webofknowlege.com), Scopus (scopus.com) and Google Scholar (scholar.google.com) yield almost no results when searching for terms like "Interac e-Transfer", "Email money transfer", "CertaPay", "Acxsys" etc. Moreover, the rare results are often based on only brief references to the system which are irrelevant for this paper, e.g. the work by Moslehpour and Azhar [11, pp. 46–47] and Lacoursière [12]. Other results – partly found by additionally using general purpose search engines – include a journal article by Ramos [13] which is little more than an advertising press release as well as surveys and reports which show the importance of Interac e-Transfers as Canadian payment method but do not consider its security or privacy either [1], [14]. Tompkins and Olivares present a qualitative analysis of "Clearing and Settlement Systems from around the world" [15] which also includes the Interac e-Transfer platform. Yet, the work's focus is on certain features of the systems, as e.g. fees and value limits, so that it mentions security aspects only in passing while privacy is not discussed at all.

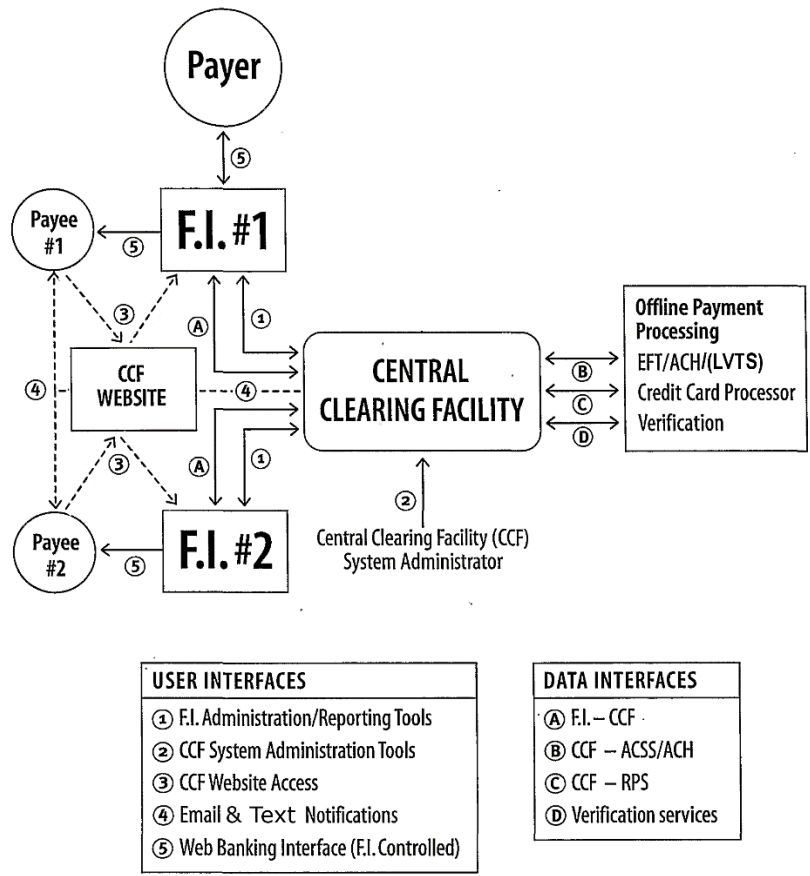

*Figure 2: Interfaces of Interac e-Transfer platform architecture*, updated version, based on original taken from *[9, Fig. 3]*

Other work by Sahut [16] examined the security and privacy of PayPal. Preibusch et al. [17] researched the leakage of purchase details from merchants to PayPal. These publications also inspired this paper but are not 1-to-1 transferable to the Interac e-Transfer platform. E.g. PayPal asks fund recipients with non-registered e-mail addresses to register them,[1] instead of authenticating the recipients via security

---

[1]  Example data included in experimental dataset (folder "Experimental Data\Support Data\PayPal Transaction")

questions. Another P2P payment platform, called *Cash App* (formerly *Square Cash*), which also allows for sending money via emails, was not the subject of a published research regarding its security and privacy to the best of our knowledge, either. Moreover, its authentication method of new customers is similar to PayPal's method of authentication and therefore disparate to the method of Interac e-Transfers [18], [19]. A certain feature of *Cash App,* called *$cashtag*, however, is similar to an improvement suggested in section 3.7.

Thus, this paper, to the best of our knowledge, is the first to analyse the security and privacy of Interac e-Transfers and to propose improvements.

The potential value of this contribution is emphasized by the existence of a pending patent assigned to the *Royal Bank of Canada (RBC)*. It explicitly refers to the Interac e-Transfer platform when stating that emails are by default an insecure method of communication as they can be intercepted and forged. It also contains the criticism that security questions may in certain cases be a weak verification method [20, p. 2]. One of the key ideas of the patent is therefore the encryption of the security response within the security question by means of public-key cryptography [20, Fig. 4]. This approach is very promising at first sight, but it only addresses certain parts of the problem as, for example, it does not consider privacy issues.

## 3. On the Security and Privacy of Interac e-Transfers

### 3.1. Scope and Boundaries of This Project

The previous sections show that the Interac e-Transfer platform is a complex system with more than 250 participants, several interfaces and types of services. Therefore, this paper focuses on certain parts and services of the system. Additionally, we faced certain constraints (time, funds etc.) in preparing this paper which limited the extent of the research (see Appendix C). The scope is defined as follows:

- The paper considers the part of the Interac e-Transfer platform which is based on internet and mobile services, especially the notification messages being sent via email and SMS. It also considers certain aspects of the financial institutes' website user interfaces. Backend parts of the system, such as the LVTS as well as the interfaces between the financial institutions and the Interac Corp. are not considered.
- International Interac e-Transfers [21] and special features for businesses [22] are out of this paper's scope.
- The paper does not consider possible legal requirements, e.g. regarding the information contained in notifications.
- The cryptographic security of e.g. *Transport Layer Security (TLS)* or *Secure Sockets Layer (SSL)* connections[2] to the financial institutes' website user interfaces or TLS connections for email delivery are considered to be a given requirement. An exception to this is discussed in section 3.2.

### 3.2. Prerequisite Observations and Related Assumptions

As shown before, the Interac e-Transfer platform uses emails and SMS to notify the customers involved in a transaction about its status. In the case of the *Standard e-Transfer* and the *Request Money e-Transfer* the messages additionally contain custom links which enable the recipient to deposit the funds after correctly answering the security question respectively to authorize the requested transfer. The information and links stay secure and confidential if a third party (adversary) is unable to eavesdrop onto the communication, i.e. is unable to read the plaintext content of email and SMS. *TLS* encryption between email servers ensures that emails cannot be eavesdropped in transit. In many cases, however, emails can likely still be eavesdropped for various reasons:

---

[2] See https://www.cloudflare.com/learning/ssl/transport-layer-security-tls/.

- Although the *Google Transparency Report* suggests that the general percentage of TLS encrypted email transfers significantly increased over the past five years [23][3] this does not apply to all email providers, apparently especially not when it is not part of their core business:
- As can be seen from Appendix D, the Canadian internet providers Bell/Sympatico, Cogeco, and Videotron do not support TLS encryption for incoming emails to their respective webmail services for customers (emails ending in bell.net, sympatico.ca, cogeco.ca and videotron.ca).[4] That means that emails sent to these addresses are transferred in plain text unless they are encrypted via end-to-end encryption such as OpenPGP[5]. It has to be mentioned that it is possible that TLS is specifically configured for transfers between the Interac e-Transfer email servers and the internet providers' respective mail servers. However, this is unlikely since enabling TLS for all incoming email transfers requires less effort than specifically configuring it for individual incoming connections (sending email servers). And even if it was the case, it would only shift the problem:
- Whenever emails are received at a target device, be it desktop computers, mobile phones, tablets, unsecured mail servers or others, they are readable in plain text unless they are encrypted via end-to-end encryption, which the notification emails for Interac e-Transfers are not. That means that malicious software running on the target device and having access to emails stored on that device can read the contents of the emails [24]. This is important to consider as therefore even emails which are sent to an email provider supporting TLS will be potentially readable by malicious software on the target device. The same applies for SMS messages which are encrypted in transit but also potentially readable by malicious software on the target device [25].
- At last, certain sources also indicate that intelligence agencies such as the NSA might have means to also eavesdrop onto TLS encrypted connections [26], [27, Secs. 2 & 5]. Of course, this is an extreme case but nevertheless relevant as section 3.3 (Experiment-Based Privacy Analysis) will show.

Therefore, this paper considers all information which is directly contained in Interac e-Transfer related notifications (SMS/email) or linked therein (web page without login) as *potentially observable*.

Based on these observations the remainder of this paper assumes different types of notification readers, respectively eavesdroppers (adversaries), which also have different levels of strength:

1. Intended recipients and senders: the intended recipients and senders are considered, too, because they might also learn unintentionally disclosed information about the other party.
2. Unintended recipients/observers: might receive emails when wrong email/mobile phone number is used or might observe emails as part of their work (e.g. system administrator).
3. Criminal individuals: might, for example, target an unsecured email server or several devices of end consumers using malicious software.
4. Criminal organizations/groups: might, for example, target several email servers or many target devices with malicious software.
5. Intelligence agencies: might target badly and well secured network devices (internet backend) to eavesdrop on unencrypted or encrypted email transfers – with different levels of probability, effort and success. Can also utilize the same methods as criminal organizations/groups.

---

[3] Report included in experimental dataset (folder "Experimental Data\Support Data\ 2019-04-02 Encryption in Transit (Google Transparency Report).pdf")

[4] At least some of the providers support TLS for downloading/synchronizing the emails but this is not considered here. Instead this refers to receiving mails from a third-party mail server.

[5] https://www.openpgp.org/

### 3.3. Experiment-Based Privacy Analysis
### 3.3.1. Definition of Privacy

Asokan et al. [28], [29] defined that a common security requirement to an electronic payment system is to ensure *confidentiality,* i.e. to limit the visibility of transaction-related information to the parties involved in the transaction. Tsiakis and Sthephanides [30] similarly claim that an electronic payment system must meet the requirement of privacy, i.e. information may not be visible to unauthorized people. Although the second definition is less precise, it adds an important aspect to the concept of confidentiality by referring to *unauthorized people* instead of *parties not involved in the transaction*: The difference is that parties involved in the transaction might also not be authorized to see certain information about the transaction. E.g. the financial institute of the sender does not need to be known by the recipient.

We define a very restrictive understanding of privacy via combining above definitions and refine it by introducing the term *necessary minimum*:

*An electronic payment system has to ensure privacy by limiting the visibility of transaction-related information to authorized parties involved in the transaction to the respective necessary minimum.*

The necessity can hereby be given by technical, legal and other reasons (for example, as the possible necessity to be able to associate a payment to a payer and therefore showing the payer's name – though this name might be a pseudonym).

### 3.3.2. Experimental Setup for Privacy Analysis

To answer research questions 4 and 5 this paper seeks to identify the Interac e-Transfer related pieces of information which are visible to the parties involved in an e-Transfer as well as to potential eavesdroppers (as described in section 3.2). Accordingly, we perform a series of 22 Interac e-Transfers between bank accounts at *The Bank of Nova Scotia (Scotiabank)*, *Royal Bank of Canada (RBC),* and *Bank of Montreal (BMO)*. This includes 12 *Standard e-Transfers,* 6 *Autodeposit e-Transfers* and 4 *Request Money e-Transfers*. The series is designed in a way to cover potential different influential factors (*type of transaction, amount of transferred money, financial institutes of sender and recipient, TLS encryption, email/SMS, first and subsequent transactions)* in as few transactions as possible due to the limitations discussed in section 3.1. In order to avoid influences from using known mail addresses we use email aliases specifically created for this experiment and an unknown mobile phone number. To identify the influence of using known addresses and mobile phone numbers, however, we in general use each email address and number twice.[6] To examine if Interac Corp. sends emails with different content to recipients not supporting TLS (e.g. with less information), we partially use aliases from an email forwarding service which, as we verified, does not support TLS for incoming emails (see Appendix E).

### 3.3.3. Results of Privacy Experiment

In order to have a scientific framework for a comparable measurement of private information – as it is contained and linked in e-Transfer-related notifications – Wagner and Eckhoff [31] compared different *Technical Privacy Metrics*. E.g. it would be possible to simply count the contained information items to calculate the "Amount of Leaked Information" [31, p. 57:15]. However, as also pointed out in [31, p. 57:15], this would not consider the individual sensitivity of each information. At the same time, it does not seem to be helpful to apply more complex private information measurement metrics, such as described in Wagner and Eckhoff (2018), since the e-Transfer platform will not be compared with other systems here. Instead, we will discuss general findings regarding the potentially observable Interac e-Transfer related information and which information would be interesting from the perspective of different types of adversaries.

---

[6] With two exceptions as can be seen in Appendix G

The privacy experiment shows that e-Transfer related notifications and web pages (linked in the notifications and not requiring a login) in general contain many pieces of private information including financial information. The transmitted information varies the most depending on the type of e-Transfer as, for example, some information only exists for certain types of e-Transfers.[7] E-Transfers which are sent to mobile phones (*Standard or Money Request*) contain slightly less information, e.g. no recipient name for Standard e-Transfers. Additionally, it can be observed that subsequent (2nd, 3rd etc.) *Standard e-Transfers* and *Money Request e-Transfers* contain a link to the recipient's *preferred financial institute*, i.e. the financial institution on whose website the recipient logged in after receiving the previous e-transfer (request) to this e-mail address. As expected by the authors, neither the support of TLS encryption at the recipient's email server nor the amount of money transferred has influence on the type and amount of observable information. That is, there is no mechanism implemented which reduces the detail or amount of information in notifications sent to non-TLS-supporting email servers or in notifications about larger amounts of money.

The experiment's results show that some information varies slightly in its concrete form. The e-Transfer senders' and recipients' name format (custom/legal/both) depends on their respective financial institute. This is obviously due to different implementations of the XML based interface to the Interac e-Transfer platform, i.e. it depends on the information which is passed on by the financial institutions via the interface. Section 3.6 discusses some problems related to this finding.

Another observation is that not all e-Transfer related confirmations (though notifications) must be sent via email or SMS. The Royal Bank of Canada enables their customers to view some e-Transfer related confirmations within a private mailbox on their website.

A last general observation is that e-Transfer related notifications are always sent within 30 minutes or less which allows conclusions to be drawn about the bank time habits of the person being eavesdropped on.

The following considers the observable information from the perspective of each potential observer.

1.  Intended recipients and senders in general can only see the information they need to know or already knew. However, two important exceptions are that the sender can see the recipient's legal name when sending an *Autodeposit e-Transfer* and that the recipient can see the sender's legal name when receiving a *Request Money e-Transfer*. Therefore, these types of transfers do not allow for anonymously/pseudonymously sending/requesting money (which *a Standard e-Transfer* in general allows, depending on the sender's financial institution name format). Additionally, the sender's financial institution is also unnecessarily revealed to the recipient.

2.  Unintended recipients/observers (e.g. system administrator, roommate, spouse) can learn information which in the best case might be embarrassing for the sender or receiver ("Why did you pay x for/to y?"), but for example they might also learn information which would allow for blackmailing the intended sender or recipient.

3.  Criminal individuals can use compromising information about previous money transfers for blackmail. They can also search emails (e.g. on a server) for potentially interesting money transfers in order to redirect them (see section 3.4). Observing the banking time habits of the victim can help in doing so.

4.  Criminal organizations/groups can do the same as criminal individuals but e.g. might have access to several email servers or many target devices. Additionally, they can use the information to forge customized phishing emails, as the observable information amongst other things contains information about people to/from whom money has been transferred, including names, financial institutions, amounts, preferred language etc.

---

[7] A recipient financial institute e.g. can only be known beforehand for an *Autodeposit e-Transfer* since the recipient of a *Standard e-Transfer* or *Money Request e-Transfer* can choose the financial institution to/from which the funds shall be transferred.

5. Intelligence agencies can use the observable information to track money flows (including personal notification/confirmation messages regarding the purposes of transfers), e.g. to identify suspects or to observe their funding and expenses.

Table 1: Potentially observable e-Transfer-related information on sender side (e-Transfer initiator)

| Type of Transfer | Information source | Overall Information Observable on Sender Side without login (Differences highlighted in bold) |
|---|---|---|
| Standard transfer to email or mobile phone | Email confirmation | Preferred language, transfer status (accepted), recipient name (custom), sender name (sender FI-specific format: custom/legal), amount of transferred money, custom notification message from sender, **custom confirmation message from recipient (if supported by recipient's FI)**, financial institute of sender |
| Autodeposit | Email confirmation | Preferred language, transfer status (accepted), recipient name (**legal** and - depending on recipient FI - also custom), sender name (sender FI-specific format: custom/legal), amount of transferred money, custom notification message from sender, financial institute of sender, **reference number** |
| Money Request | Email confirmation | Preferred language, transfer status (accepted), recipient name (custom), sender name (sender FI-specific format: custom/legal), amount of transferred money, custom notification message from sender, **custom confirmation message from recipient**, financial institute of sender |

The tabular summary of the observable information is shown in Table 1 (above) and Table 2 which can be found in Appendix F. The detailed tabular representation of observable data is shown in Appendix G.

### 3.3.4. Probable Reasons for Privacy Problems

Since the above clearly suggests that the e-Transfer related information can be unintentionally or purposefully misused for different unauthorized activities, the question arises why it is being sent at all? Based on our observation we tentatively conclude the following:

1. The process of the *Standard e-Transfer* and *Request Money e-Transfer* requires the recipient to perform an action (depositing the funds or authorizing to send funds).
2. This action requires the recipient to login to a financial institution.
3. Within the e-Transfer platform it is technically necessary to provide (and use) a custom link contained within the e-Transfer notifications to perform these actions. This creates the danger of phishing mails which mimic this type of email as can be seen from Appendix H. It can be assumed that the personal information contained in the notifications therefore has the purpose to show its legitimacy so that the recipient knows that the link is safe (and not a phishing link). Since this is achieved by showing secret knowledge about the transfer, we call this approach *exposed knowledge verification*.
4. Additionally, it can be assumed that including detailed information about the transfers is also convenient for the customers since they can track the status of e-Transfers without logging in to their financial institution's website. This would explain why detailed information is also contained in confirmations which do not contain the aforementioned custom links.

Section 3.7 will propose improvements to address these problems, namely the requirement of the custom link and the *exposed knowledge verification* as well as the convenient tracking of e-Transfer statuses.

### 3.4. Possible Attack Scenario Arising from Identified Privacy Problems:

The previous section shows that the information contained in Interac e-Transfer-related notifications can be misused for fraudulent and other criminal activities. In the following, a possible attack that could for example be carried out by a criminal group is described in more detail.

1. The group commits financial identity theft, as e.g. described by Aïmeur and Schönfeld [32], by hijacking an individual's Canadian bank account. This might be easier than expected since e.g. the banks considered in this paper (BMO, RBC, Scotiabank) rely on security questions for verifying new browsers (see Appendix I). Security questions, however, are shown to be insecure by Schechter, Brush and Egelman [33].

2. The group eavesdrops on a larger number of email recipients, e.g. by compromising an email server, or many end user devices, specifically searching for *Standard Interac e-Transfers* of larger amounts of money. By observing e-Transfer related notifications it can also observe the usual banking times of its victims in order to identify a good time for trying to redirect a money transfer. I.e. the attackers have more time to figure out the answer to the security question at a time when the victim usually is not banking online.

3. They now open the deposit link in the e-Transfer notification and login to the hijacked bank account to see the security question. Depending on its kind and strength (see section 3.5.2), they can try to guess it or to research the answer in previous emails or social networks. Alternatively, they might simply log the answer when the *legitimate* recipient answers the security question. In the case of a recurring payment with the same security question they know the answer for the next transfer.

4. When the group knows (see step 3 above) or correctly guesses the answer to the security question it can deposit the money in the hijacked account.

5. The deposited funds in the hijacked account can then be used for buying goods online or they can be forwarded to an international account. As long as the group quickly rebalances the funds in the account, the legitimate owner of the account might not even notice that activity. The sender of the e-Transfer will not notice the redirection (see section 3.6) and the original *legitimate* recipient will have to contact the bank/Interac to learn where the funds were deposited respectively to report the fraudulent activity.

6. The attacker group can continue this attack until the hijacked account is frozen. However, since the probability of detection will increase over time it can also decide to clear the account e.g. when reaching the daily maximum of e-Transfer deposits, which is usually CAD 10,000. Subsequently they can continue their attack with another hijacked account. Thus, they have increased their 'profit' by e.g. CAD 10,000 in comparison to only clearing the account (without using it for redirecting e-Transfers).

### 3.5. Experiment-Based Security Analysis

### 3.5.1. Definition of Security

Asokan *et al.* [28], [29] defined that besides *confidentiality* another common security requirement to an electronic payment system is to ensure *integrity and authorization,* i.e. – from an individual's perspective – that no funds are transferred from that individual's account without his/her authorization. The same can apply for receiving funds, e.g. to be able to reject unsolicited bribes [28], [29]. Other security requirements are discussed by Sahut [16], Asokan *et al.* [28], [29] and Tsiakis and Sthephanides [30] but are out of this paper's scope. However, the next sections will show that improving the security of the Interac e-Transfer platform necessitates the definition of additional requirements (see section 3.7).

### 3.5.2. On the (In-)Security of Security Questions

Research questions 1 and 2 consider the possibility of redirecting Interac e-Transfers (1) and the (in-)security of Interac e-Transfer security questions (2).

As described before, it is possible to see a *Standard e-Transfer's* security question when selecting the deposit link in an – e.g. eavesdropped – notification and logging in to the website of a financial institution. I.e. the security is not provided by withholding the security question but by having to know the correct answer. We assume that it will often be possible to guess or research the correct answer e.g. by searching previous mails which the sender sent to the recipient. RBC's pending patent for enhancing e-Transfer security similarly states that the "verification process can be weak if the challenge [...] is based on information that [...] may be discovered by a third party through a social media website or otherwise" [20, p. 2]. Additionally, security questions have been researched before: Schechter et al. [33] came to the overall conclusion that the "security of personal questions appears significantly weaker than passwords" [33, p. 386]. Questions which were created by the participants of their survey in fact could be guessed in about half of the cases by either acquaintances who accompanied them or even by complete foreigners [33, p. 386].

Other authors also pointed out that the usage of security question often creates privacy problems, too: if actual private information is used for specifying security questions/answers, this information might additionally become public in the case of a security breach.

It is important to note that Schechter, Brush and Egelman [33] examined security questions which would be used e.g. to verify someone's identity before resetting that someone's password. That means that their study considered security questions that did not serve to verify the legitimacy of a second individual – who also knows the answer to the security question – but only to verify the first individual itself. However, we argue that a security question and answer shared by two persons can only be as strong (or weak) or even weaker than any security question the two persons might create for authenticating themselves. This assumption is based on set theory: the set of security questions and answers which a person would want to share with a second person – or which the second person already knows – can only be a subset of both persons' *personal security questions*, respectively the intersection of these sets. More specifically, it will be the intersection of the two sets of *previous knowledge-based security questions* and the two sets of – for this purpose – *made up security questions* as shown in Figure 3. When financial institutions like the *Canadian Imperial Bank of Commerce (CIBC)* provide pre-defined security questions, see [34], the set of predominantly used questions might be even more limited. We think it would be interesting to empirically research these problems – especially regarding their potential implications for the security of *Standard e-Transfers* – but, like Schechter, Brush and Egelman [33, p. 386], we prefer to suggest using an alternative authentication method, or to use an implicit authentication as described in section 3.7.

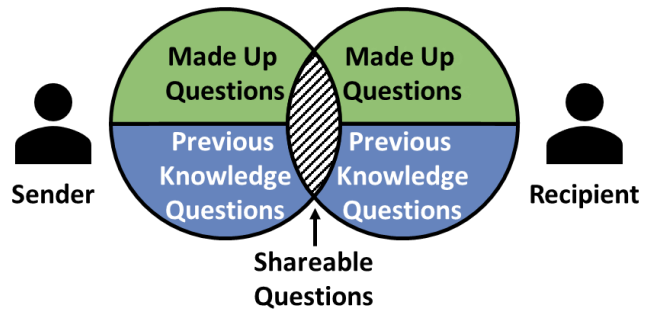

*Figure 3: Set of security questions that are shareable between sender and recipient*

### 3.5.3. Experimental Setup for Security Analysis

The set up for the security experiment is similar to the privacy experiment in that we use the author's three bank accounts at BMO, RBC and Scotiabank to perform a series of 9 *Standard e-Transfers*. For this experiment, however, we always use the author's legitimate email addresses as sender addresses (instead of generic email addresses created for the privacy experiment). The *legitimate recipients'* names and email addresses are made up from common Canadian given names and last names. For a certain series of transfers, the *legitimate recipient has* the name of one of the authors (*first author*) and a specifically created email address which contains this author's name. Additionally, we use a mobile phone number that the author sending the related e-Transfers has not used before. All recipients' email addresses were registered at an email forwarding service that does not support TLS for incoming mails.

The other important difference is that all transfers – except for one[8] – are not deposited by these invented recipients – or the *first author* – but instead by the *second author (into the second bank account owned by him)* who hereby also takes the role of the attacker. Only one transfer is deposited by the *first author* in order to suggest that this is his new "legitimate" email address – from the perspective of a fraud scoring engine. We simulated the eavesdropping of the attacker by simply forwarding the email and SMS notifications to him.

Overall, we take care to correctly simulate the fraudulent activity: For all transfers, the *first* and the *second author* are at different locations and use different IP addresses. For depositing the transfers, the *second author* uses a device and browser which he previously did not use. He also uses a third IP address (by mobile phone tethering) in order to simulate that an adversary hijacked his account.[9] The series of e-Transfers covers different sender accounts and money amounts (CAD 0.10 to CAD 1,900). It covers different numbers of failed attempts to correctly answer (what we consider to be) a weak security question. During a sub-series of three transfers with a more complex security question/answer the *first author* deliberately failed to answer correctly even in the fourth (last possible) attempt. After that, the sum of funds in these transfers is resent in a new transfer and then successfully redirected. This seeks to simulate an adversary who tricked the sender to resend the funds with the same question after having researched the correct security answer. The experimental setup is shown in Figure 4.

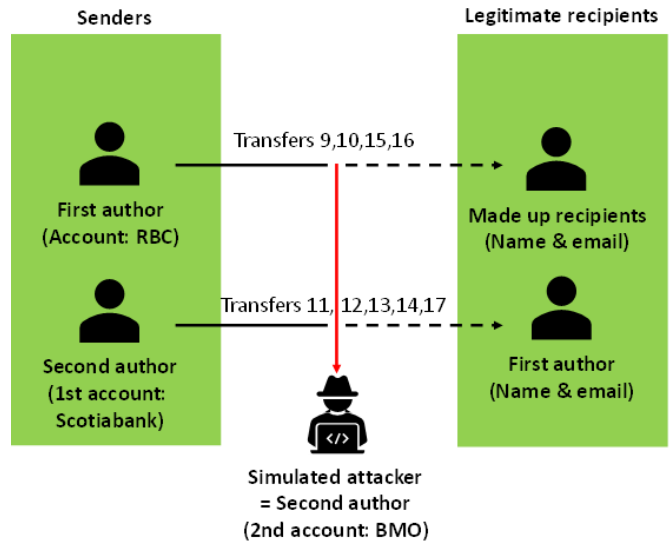

*Figure 4: Experimental setup of security experiment*

### 3.5.4. Results of Security Experiment

Overall, the results of the security experiment confirm the authors' early observation – that it is possible to redirect Standard Interac e-Transfers to a third-party's bank account. Additionally, the results show that the amount of money being transferred or possible knowledge about previous deposits of the recipient do not influence this property, i.e. redirects are also possible for bigger amounts of money (verified for up to CAD 1,900) and also for email addresses which have been used to deposit funds into another – legitimate – person's bank account before. It is also possible to redirect funds which have been sent to a mobile phone number, though we did not verify if it is possible to do that after previously depositing a transfer into the legitimate account (due to a limitation of available mobile numbers[10]).

As described above, the experiment sought to also simulate more suspicious behaviour/patterns, e.g. by looking at security questions, failing to answer them, etc. Yet, this also did not influence the successful deposit into the simulated hijacked bank account. The detailed results of the experiment are presented in tabular format in Appendix J.

---

[8] This transfer is deposited by the first author in order to suggest that this is his new "legitimate" email address – from the perspective of a fraud scoring engine.

[9] Consequently he had to answer a security question when logging into his bank account for the first time from this device.

[10] The mobile number used in this experiment belonged to the first author who therefore would have had to send funds to his own account from which the transfer was initiated – which is not possible.

We additionally noticed that the sender of an e-transfer will not know that it has been redirected, since the confirmation message to the sender always contains the recipient name which the sender originally specified. The legitimate recipient will of course eventually notice that the transfer has been redirected because opening the deposit link shows that the transfer has already been deposited. He might also notice that the default link for the preferred financial institute changed when another transfer is sent to the same email address and the attacker's financial institution is different from his financial institution.[11]

However – as will be discussed in detail in sections 3.5.5 and 5 – one has to take into consideration that the *redirectability* might be an *intended or tolerated undocumented feature* and/or that the limitations of our experimental setup influenced the results.

### 3.5.5. Reasons for Security Problems

In our opinion, the identified possibility to redirect Standard Interac e-Transfers and the described attack scenario is due to several platform-specific design decisions/properties.

1.  Standard e-Transfers are not strictly associated with the legitimate recipient.
2.  The platform mainly assumes that email/SMS are neither eavesdropped nor sent to unintended recipients, because in this case it is impossible to identify transfers of bigger funds, to open the deposit link, or to learn other sensitive information.
3.  In the case that an eavesdropper/unintended recipient can read a notification, the platform relies on the security question being strong enough to keep them from depositing the funds.
4.  The fraud scoring engine/proprietary risk management used by Interac Corp. [6], [7] fails to identify the redirection of a transfer to an illegitimate recipient, e.g. based on the recipient's name.
5.  At least some accounts at banks, who are participants in the e-Transfer platform, seem to be weakly secured against financial identity theft (*hijacking* of account), e.g. due to the usage of security questions.

As previously mentioned, it is possible, however, that it is intended that consumers are able to redirect/forward e-transfers to other recipients: to the best of our knowledge, there is no policy stating that the name of the recipient must match the name which the sender specified for the recipient. It can be assumed that this would create issues e.g. seeing that senders might use nicknames or other name variations (including typos) when specifying the recipient name. Additionally, it is clearly intended that a recipient may choose into which legitimate account a transfer shall be deposited.

We are however sceptical that the *redirectability/forwardability* to another person is an intended feature since we did not find any related information.

However, if it is a tolerated property or even an undocumented feature, it – as the experiment shows – can be misused for fraudulently redirecting transfers which should obviously be avoided. We therefore propose improvements to overcome this issue – and other issues – in section 3.7.

### 3.6. Additional Observations

During the execution of our experiments and the analysis of the results, we noticed further peculiarities and flaws regarding the Interac e-Transfer platform, also regarding the Autodeposit and Request Money e-Transfers. The security relevant observations and our considerations are discussed below and are taken into account for proposed improvements in section 3.7. A discussion of the other observations is out of scope of this paper, but we have documented these and our first considerations in this regard in Appendix K.

---

[11]  This is not applicable for SMS or when the account to which the transfer was redirected belonged to the same financial institute.

1. The E-Transfer sender email addresses at Scotiabank and RBC are not verified before they can be used. Therefore, notifications can get lost or sent to unintended recipients when incorrect email addresses are specified; similarly request notifications can show incorrect sender email addresses.
2. The initiation of e-Transfers at BMO, Scotiabank, and RBC do not have to be authorized individually e.g. using a second authentication method such as a *transaction authentication number (TAN)*[12]; the login authentication is sufficient. Authorizing each transfer would be less convenient but more secure.
3. It is impossible to reject an Autodeposit e-Transfer (at least at BMO, RBC, Scotiabank). This may lead to unsolicited bribery or to problems when someone specified the wrong recipient. Depending on the sender's financial institute, the recipient will not be able to see the sender's legal name. Additionally, the legal name can be ambiguous and the recipient might not know the sender. Since the sender email address is not shown to the recipient for this type of e-Transfer, it is difficult to contact the sender or to simply send the money back.
4. If a recipient has configured the Autodeposit feature for a certain email address, the sender will be shown the recipient's legal name (and depending on the recipient's financial institution also the profile name) after entering the target email address in the web interface. This is obviously in order to verify that the correct recipient has been specified, which, however, is very problematic:
   a. It is possible that showing the name is not sufficient but rather misleading, e.g. if the sender is not sure about the recipient's name and guesses that e.g. *Michael Miller* has the email address *michaelmiller@domain.com* although it actually was *m.miller@domain.com*. If a recipient with that email address (who registered that address for Autodeposit) exists, it is very likely that showing the legal name will misleadingly suggest that the correct recipient's address has been specified.
   b. Showing legal names of Autodeposit recipients also does not allow for anonymously/pseudonymously receiving transfers via this e-transfer type.
   c. The interface might be misused for retrieving the legal name of a person of whom someone only knows the email address. Additionally, random email addresses can simply be tested e.g. to create tailored phishing attacks (using correct legal names). For testing, hijacked bank accounts could be used. We noticed that the interface at BMO, for example, allows for testing email addresses in quick succession but we did not test the limits of that. This problem might be researched in future work – e.g. with experiments or examining the Interac e-Transfer API[13].

## 3.7. Proposed Improvements

The identified privacy and security issues – not all of them related to the redirection of transfers – can be overcome in different ways. Approaches include short-term implemenTable recommendations for Interac e-Transfer users and middle to long-term implemenTable proposals to financial institutions, email providers, and Interac Corp. These approaches, however, do not address all problems so that we just outline them in Appendix M.

In contrast to this, as a first step, we propose additional requirements/properties that must be met by the Interac e-Transfer platform to overcome the identified privacy and security problems. That is because these problems originally result from certain properties and design decisions regarding the Interac e-Transfer platform as described in sections 3.3.4, 3.5.5 and 3.6. We propose the following requirements (here including explanations referring to the Interac e-Transfer platform).

---

[12] See https://www.investopedia.com/terms/t/transaction-authentication-number-tan.asp.
[13] See https://developer.interac.ca/login?redirect=https://developer.interac.ca/e-transfer/.

1. *Intrasystem security and privacy*: The security and the privacy of the system must not rely on external systems (such as SMS or email) which are not under the service provider's (here Interac Corp.) control. Accordingly, security and privacy must also not be based on a human component (such as a security question). Where the human factor has to come into play (e.g. when creating passwords), technological means must minimize the risk of unsafe system configuration and usage.
2. *Transfer-directability*: E-Transfers of all types have to be technically linked to a certain recipient, in order to ensure that it cannot be redirected to another recipient. The recipient may nevertheless have the option to choose a certain bank account.
3. *Identifier verification*: The sender identifier (currently the notification email address) for all types of e-Transfers must be verified to belong to the person who wishes to use them before first use. This is in order to avoid accidental or intended specification of wrong identifiers.
4. *Technical recipient identifier-verifiability*: The recipient identifier (currently the target email address or phone number) for all types of e-Transfers must be verifiable regarding accidental misspecification. The verification should be based on a technical verification that does not expose personal information. At the same time, it should not expose information regarding whether a certain identifier does not exist or if it was incorrectly specified in order to avoid snooping of identifiers.
5. *Transfer rejectability/returnability*: Recipients of unsolicited e-Transfers must be able to either reject the transfer or – in the case of an Autodeposit – to return the transfer without needing to know the sender's identity.
6. *Individual e-Transfer initiation, fund-withdrawal and device change authorization*: An e-Transfer initiation, the withdrawal of funds from an account (via e-Transfer request) and device changes always have to be individually authorized with a one-time (password) method.
7. *Minimum information disclosure*: Notifications to e-Transfer customers sent via an unsecure channel must not contain any sensitive personal/financial information. Notifications sent via secure channels must not reveal unnecessary information, including banking times of e-Transfer users, to other users or third parties.

Based on the proposed additional requirements above, we propose several related improvements to the Interac e-Transfer platform that holistically address all identified problems and meet the desired properties as briefly discussed in the next section: The Standard e-Transfer should not be used in the future but replaced by a new type of e-Transfer, which we call *Directed e-Transfer*: This type of transfer is strictly associated with a certain customer who nevertheless may choose at which financial institution (and into which account) a transfer shall be deposited or if the transfer shall be rejected. In order to be able to achieve this, each e-Transfer customer has to manually register the bank accounts which shall be available for selection as described below. Alternatively, the accounts held by a certain customer might automatically be mapped to him or her (and made available for selection) based on an already known unique identifier such as the social security number (given that this kind of identifier is already known to Interac).

1. The identifier, to which the transfers can be sent, is either a personal email address (as before) or, alternatively, a freely selecTable but unique alphanumeric string with a certain maximum length (e.g. 10 characters). Mobile phone numbers may not be used as identifiers since they are more likely passed on to foreigners if e.g. a mobile phone contract is ended. The identifier can be selected and registered after logging into a financial institution's website. The registration requires the specification of a notification email address or mobile phone number which may only be linked to one identifier at any time. Target identifiers based on email addresses must be registered and verified as well. After registration, both types of target identifiers are called *InteracID*. Technically they are the same, they differ only in length and regarding the existence of the @-character in email-based *InteracIDs*.

Additionally, the notification email address of an email-based InteracID is the email on which it is based. The InteracID of the sender is not revealed to the recipient.

2. Each *InteracID (*including email-based InteracIDs), is assigned a random 3-digit *security code* which must be entered correctly in addition to the InteracID when sending an e-Transfer to this identifier.

3. Money Request e-Transfers and Autodeposit e-Transfers are still available but must also be sent to InteracIDs, including the specification of the *security code*. That means that InteracIDs used for Autodeposits are associated with only one bank account and that in this case they provide similar functionality as do the email addresses which are currently registered for Autodeposit. Additionally, it also means that a person can own several InteracIDs (e.g. one InteracID which allows for choosing between accounts and one or more additional account specific InteracID for Autodeposits).

4. The recipients of an e-Transfer Autodeposit may choose to return the money to the sender without learning the identity of the sender – unless the sender specified his real name. This is technically maintained by keeping track of previous transfers which can be reversed by the user (e.g. within a certain time frame). The InteracID of the sender hereby must not be revealed to the recipient.

5. The initiation of any type of e-Transfer, sending funds to meet an e-Transfer request and device changes have to be authorized via state-of-the-art two-factor authentication (e.g. password and TAN generated in banking app).

6. Customers are notified about e-Transfers by generic messages stating that they have received an e-Transfer of a certain type (Directed, Autodeposit or Request) without giving further details and which asks them to login to their financial institution's website. Customers can opt-in (or opt-out) regarding that these emails are cryptographically signed. However, a login link is not provided within the notification. The customer knows to which InteracID the e-Transfer was sent, based on the email address/mobile phone number at which the notification was received. Accordingly, the customer also knows to which financial institution he has to login, whereby he may choose from any account (i.e. financial institution) which is assigned to the specific InteracID. Additionally, customers can be contacted via secure communication channels, e.g. within the respective financial institution's mobile banking application and/or an application to be developed and provided by Interac Corp to notify them about an e-Transfer, including status updates. However, only minimum necessary information is contained in the notifications. Accordingly, users may choose if a status update/notification about an initiated transfer shall be communicated to the other party (sender/recipient) in real time or not.

## 4. Evaluation

The main contribution of this paper is the research into potential security and privacy issues related to the Interac e-Transfer platform and the proposal of additional requirements and improvements to address the identified issues.

We designed our experiments in a way that sought to avoid the results being influenced or biased by factors like the usage of known e-mail addresses, size of transferred funds, TLS support, sender and recipient financial institutions, usage of email or SMS, and first-time/subsequent transfer. Additionally, we verified or supported our assumptions regarding the potential observability of information by either technical means (e.g. regarding TLS support) or by reviewing relevant sources. Thus, we are confident that our privacy-related results are correct and in general also apply to other financial institutions participating in the Interac e-Transfer platform. Regarding our security related results, we consider them to be tentative findings, due to the given limitations in the experimental setup. For example, the accounts we used have at least some relation to each other: they are owned by the same person and/or have exchanged funds via legitimate transfers before (e.g. during the privacy experiment). This might have influenced the security related results, e.g. considering the possibility that Interac Corp. automatically detected suspicious behaviour in the accounts but did not take action after manually reviewing the matter. However, we consider this to be unlikely since we carefully sought to simulate real fraudulent behaviour. Thus, we would have expected Interac Corp. or our financial institutions to at least contact us in the case of its detection. Additionally, we did not notice any delayed transactions (taking more than 30 minutes). A manual review would probably have taken more time.

With regards to the proposed additional requirements (and improvements), we argue that it is possible to tentatively evaluate their validity by mapping the identified problems to requirements and improvements as presented in Appendix N. Future research can verify the approach by applying theoretical proof (see section 5).

## 5. Discussion, Conclusion & Future Research

This paper has identified and addressed problems regarding the privacy and security of the Interac e-Transfer platform and thereby closed the gap created by the absence of earlier examinations into that platform. Our results show that under realistic assumptions – regarding the abilities of todays' eavesdroppers'– the platform fails to protect its customers' privacy due to design decisions which may well have been state-of-the-art when the platform came into being. Nowadays, assuming the privacy and security of unencrypted emails and security questions is unacceptably risky. Additionally, the simulated attack's success shows that *Standard e-Transfers* are potentially insecure against redirections given that attackers invest enough effort to hijack someone's bank account, to identify transfers of higher values, and to research or guess the answer to security questions. Although the findings regarding security should be generally verified by future research, they are already supported by a recent news article by Johnson [35]. Future researchers will probably have to obtain the consent of Interac Corp. in order to ensure that non-researcher participants of more extensive experiments are not exposed to the risk of having their bank accounts frozen. Discussing the issues with Interac Corp. was unsatisfactory from our point of view (see section 6). Consequently, we have proposed a holistic approach to address both the privacy and security problems.

Further research could also be done to evaluate and compare today's Interac e-Transfer implementations by including more participating financial institutions, e.g. to identify potential inconsistencies, flaws (e.g. regarding the Autodeposit user interface), and other usability issues such as those documented in Appendix K. It is also possible that some of the identified issues may be created or exacerbated by the respective implementations of the financial institutions.

Additionally, researching security questions as an authentication method between two parties would extend earlier research which examined security questions as an authenticating method for web logins (one party). This may support or reject our findings on the security of Standard Interac e-Transfers in general and in particular our set theory-based approach to applying these earlier results to this scenario.

Finally, we would be very happy to see other researchers evaluating our proposed improvements with theoretical approaches such as UML-based secure systems development [36] or using our suggestions to propose their own alternative approaches. Either way, we hope to see according changes in the Interac e-Transfer platform, as soon as possible.

## 6. Epilogue

On March 13th, 2019 we contacted Interac via phone in order to inform them about our ongoing research (which was at that point an early stage university course project report) to give them a chance to verify our results and/or to add additional information to the report. After a confirmation call on March 18th, 2019, we did not hear back from them due to Interac sending their answer to an incorrect email address. Also, we decided to not contact them again before submitting our initial report (as a university course project) in order to avoid issues regarding confidentiality and delays. Yet, we already discussed that we might contact Interac again in case we intend to submit the report as a paper for publication. We submitted the initial project report at the University of Ottawa on April 9th, 2019.

On May 12th, 2019 CBC News posted an article which reported on fraudulently redirected Interac e-Transfers [35].

This article confirmed some of our findings regarding the security of *Standard Interac e-Transfers*. Due to this, and since we were looking into publishing the report as a paper, we contacted Interac again and got a response stating that they would like to review the report before publication. Therefore, we incorporated minor changes in the report, e.g. an early epilogue section and some clarifications based on the feedback we got for the report and the intended submission for publication. This version of the report was then shared with Interac for review on May 23rd, 2019.

In their review of the report (June 3rd, 2019) Interac pointed out that we had no access to confidential information regarding the layers of security and fraud protection, the usage of outdated sources like e.g. a 2002 patent, and the lacking consideration of legal requirements [10]. They also emphasized that they share "important tips" [10] regarding security including the usage of different layers of security, the usage of strong passwords, not to choose easy guessable security questions, and using email providers which use TLS encryption [10]. The full text of the review can be seen in Appendix O.

We have taken the feedback into account in the version of the report presented in this paper. However, we additionally want to reply to the feedback as follows:

As pointed out in the paper, we deliberately focused on publicly observable parts of the system, such as email and SMS notifications and user interfaces. Our research regarding these parts seem to show that the privacy and security issues resulting from certain design decisions which manifest in specific contents and features of these parts are not mitigated by the mentioned layers of security and fraud protection, which are unknown to us. It is possible that certain fraudulent activities are detected by these systems and that our experiments were unsuitable to show that. Yet, the CBC News article by Johnson [35] confirms that it is possible to redirect e-Transfers, obviously similar to the activities simulated by us, to commit real Interac e-Transfer fraud.

Regarding the usage of "outdated" [10] sources, we want to stress that the 2002 Canadian patent is the only (out)dated source we use in connection with describing the Interac e-Transfer platform. All other sources for this are much more recent (2017 to 2019). Also, the status of the patent is still *pending* and it is also still active as Australian patent "AU2002227835B2", GB patent "GB2389443B" and US patent "US7844546B2". All of them have been filed in 2002 but have only been granted in 2007 (AU), 2004 (GB) and 2010 (GB). Moreover, we use this source only for describing the backend of the system, which was not part of the investigation regarding privacy and security. The more recent sources do not contradict the system as it was described in this patent; on the contrary, they seem to show that the general mechanics are still the same. Finally, the RBC patent, which directly targets to improve the security of Interac e-Transfers, has been filed in 2017 – and we argue that RBC would be aware of updated information regarding (more secure) internal workings. Instead, the patent does explicitly refer to the Interac e-Transfer platform when stating that emails are by default an insecure method of communication as they can be intercepted and forged (see section 2.4). It also criticizes that security questions may in certain cases be a weak verification method.

Furthermore, we do not think that our criticism of privacy problems is irrelevant just because we did not consider legal requirements. It is possible that certain or even all information we found must be contained in notifications and confirmations due to legal reasons. Yet we doubt this and have not received any specific information from Interac that would indicate otherwise. Moreover, even in the case that this assumption is wrong, it does not mean that the system could not be improved to overcome the privacy issues, e.g. using secure channels for communication. We also did not find any publicly available information by Interac recommending the usage of TLS enabled providers. In contrast, we want to point out that the usage of TLS would only reduce the risk of being eavesdropped since e.g. malware might be used on target devices (see section 3.2).

From our point of view, the recommendation to use strong passwords and good security questions is not sufficient either: As described in section 3.7, we claim that a secure system may not rely on the human factor without technically ensuring that the likelihood of unsafe system configuration and usage is minimized. Additionally, the use of security questions creates other problems, e.g. the possibility of eavesdropping onto earlier use of the same passwords and privacy issues in the case of data leaks, as shown in sections 3.4 and 3.5.2.

Lastly, we want to mention that Interac somehow criticized our way of communicating with them, stating in their reply:

> "At Interac, we support education studies to improve our products, services and customer experience. We engage with students on a regular basis. Generally, they come to us at the beginning with a business problem and we work with them under confidentiality agreements, as our technology and systems are proprietary." [10]

We want to reply that we contacted them before conducting our main experiments and writing the major parts of the report but that we did not get feedback due to a mistake on their side. After finishing the report, we gave them a second chance to review it, which they did. Their reply states, however, that they "are unable to disclose about how [their] technology and systems work, the layers of security and fraud protection that [they] have in place, the advanced technologies that [they] employ, and how [they] work with [their] financial institution partners" [10].

## 7. References

[1]     M. Tompkins and V. Galociova, "Canadian Payment Methods and Trends: 2018," 2018.

[2]     Interac Association/Acxsys Corporation, "Use of Interac e-Transfer service surges in 2018," *Interac Reports*, 2019. [Online]. Available: https://newsroom.interac.ca/use-of-interac-e-transfer-service-surges-in-2018/. [Accessed: 26-Mar-2019].

[3]     CertaPay Inc., "CERTAPAY AND CANADIAN BANKS LAUNCH WORLD'S FIRST REAL-TIME, BANK-TO-BANK EMAIL MONEY TRANSFERS," *CertaPay Press Release*, 26-Jun-2002. [Online]. Available: https://web.archive.org/web/20030605141306/http://certapay.com/en/newsEvents/release260602.cfm. [Accessed: 26-Mar-2019].

[4]     Interac Association/Acxsys Corporation, "Why 2018 was another banner year for Interac e-Transfer," *interac.ca*, 2019. [Online]. Available: https://newsroom.interac.ca/why-2018-was-another-banner-year-for-interac-e-transfer/. [Accessed: 26-Mar-2019].

[5]     Interac Corp., "Interac - For Consumers," *www.interac.ca*, 2018. [Online]. Available: https://www.interac.ca/en/interac-e-transfer-consumer.html. [Accessed: 29-Mar-2019].

[6]     Interac Corp., "Interac e-Transfer Security," *www.interac.ca*, 2018. [Online]. Available: https://www.interac.ca/en/interac-etransfer-security.html. [Accessed: 29-Mar-2019].

[7]     Interac Corp., "Interac e-Transfer API," *Interac Developer Centre*, 2019. [Online]. Available: https://developer.interac.ca/interac-e-transfer/. [Accessed: 29-Mar-2019].

[8]     Interac Association/Acxsys Corporation, "Principles of a Modernized Payments System," Mar. 2017.

[9]     J. Fleishman and Z. Fuerstenberg, "Online Payment Transfer And Identity Management System And Method," Canadian Patent CA 2435909, 2002.

[10]    A. Vaughan, "RE: Report on security and privacy of Interac e-Transfers - University of Ottawa Research Report," Interac Corp, 03-Jun-2019.

[11]    M. Moslehpour and A. Azhar, "IMPACT OF CUSTOMER SATISFACTION ON INTENTION TO SWITCH IN REMITTANCE COMPANIES IN TAIWAN," in *13th Asia Pacific Forum for Graduate Students Research in Tourism*, 2014, pp. 2F 45-58.

[12]    M. Lacoursière, "Analyse de la trajectoire historique de la monnaie électronique," *Les Cah. droit*, vol. 48, no. 3, pp. 373–448, 2007.

[13]    R. Ramos, "Canada launches email money transfers," *Card Technol. Today*, vol. 14, no. 9, pp. 1–2,

2002.

[14]    C. S. Henry, K. P. Huynh, and A. Welte, "2017 Methods-of-Payment Survey Report," Bank of Canada, 2018.

[15]    M. Tompkins and A. Olivares, "Clearing and Settlement Systems from Around the World: A Qualitative Analysis," *Bank Canada. Staff Discuss. Pap.*, vol. 14, no. 5, 2016.

[16]    J.-M. Sahut, "Security and Adoption of Internet Payment," in *2008 Second International Conference on Emerging Security Information, Systems and Technologies*, 2008, pp. 218–223.

[17]    S. Preibusch, T. Peetz, G. Acar, and B. Berendt, "Shopping for privacy: Purchase details leaked to PayPal," *Electron. Commer. Res. Appl.*, vol. 15, pp. 52–64, 2016.

[18]    R. Andersen, J. Bekmann, B. Grassadonia, D. Perito, and P. T. Westen, "Payment Transaction By Email," United States Patent No. US 9,165.291 B1, 2015.

[19]    N. Dieker, "Square Cash: The Free Alternative to PayPal and Venmo," *thepennyhoarder.com*, 08-Jun-2018. [Online]. Available: https://www.thepennyhoarder.com/smart-money/square-cash/. [Accessed: 01-Apr-2019].

[20]    A. T. K. Lau, E. U. Ortiz, A. Gupta, M. Sharma, L. J. Manuel, and T. J. T. Su, "System And Method For Message Recipient Verification," US Patent No. US 2018/0159865 A1, 2018.

[21]    Interac Association/Acxsys Corporation, "Interac e-Transfer goes international in collaboration with nanopay," *Interac Reports*, 29-Nov-2017. [Online]. Available: https://newsroom.interac.ca/interac-e-transfer-goes-international-in-collaboration-with-nanopay/. [Accessed: 26-Mar-2019].

[22]    Interac Corp., "Interac e-Transfer solutions for business," 2019. [Online]. Available: https://business.interac.ca/interac-e-transfer/. [Accessed: 07-Apr-2019].

[23]    Google, "Email encryption in transit," *Google Transparency Report*, 2019. [Online]. Available: https://transparencyreport.google.com/safer-email/overview?hl=en&encrypt_out=start:1385856000000;end:1554163200000;series:outbound&lu=encrypt_in&encrypt_in=start:1385769600000;end:1554163200000;series:inbound. [Accessed: 02-Apr-2019].

[24]    D. Zhang, Y. Guo, D. Guo, and G. Yu, "Privacy Leaks through Data Hijacking Attack on Mobile Systems," in *4TH ANNUAL INTERNATIONAL CONFERENCE ON INFORMATION TECHNOLOGY AND APPLICATIONS (ITA 2017)*, 2017, vol. 12.

[25]    H. Abualola, H. Alhawai, M. Kadadha, H. Otrok, and A. Mourad, "An Android-based Trojan Spyware to Study the NotificationListener Service Vulnerability," *Procedia Comput. Sci.*, vol. 83, pp. 465–471, 2016.

[26]    S. Checkoway *et al.*, "On the Practical Exploitability of Dual {EC} in {TLS} Implementations," in *23rd {USENIX} Security Symposium ({USENIX} Security 14)*, 2014, pp. 319–335.

[27]    B. Schneier, "NSA surveillance: how to stay secure | Bruce Schneier | US news | The Guardian," *www.theguardian.com*, 06-Sep-2013. [Online]. Available: https://www.theguardian.com/world/2013/sep/05/nsa-how-to-remain-secure-surveillance. [Accessed: 03-Apr-2019].

[28]    N. Asokan, P. Janson, M. Steiner, and M. Waidner, "State of the art in electronic payment systems," in *Emphasizing Distributed Systems*, vol. 53, M. V. B. T.-A. in C. Zelkowits, Ed. Elsevier, 2000, pp. 425–449.

[29]    N. Asokan, P. A. Janson, M. Steiner, and M. Waidner, "The state of the art in electronic payment systems," *Computer (Long. Beach. Calif).*, vol. 30, no. 9, pp. 28–35, 1997.

[30]    T. Tsiakis and G. Sthephanides, "The concept of security and trust in electronic payments," *Comput. Secur.*, vol. 24, no. 1, pp. 10–15, 2005.

[31]    I. Wagner and D. Eckhoff, "Technical Privacy Metrics: A Systematic Survey," *ACM Comput. Surv.*, vol. 51, no. 3, pp. 57:1--57:38, 2018.

[32]    E. Aïmeur and D. Schönfeld, "The ultimate invasion of privacy: Identity theft," in *2011 Ninth*

*Annual International Conference on Privacy, Security and Trust*, 2011, pp. 24–31.

[33]   S. Schechter, A. J. B. Brush, and S. Egelman, "It's no secret - Measuring the security and reliability of authentication via 'secret' questions," *Proc. - IEEE Symp. Secur. Priv.*, pp. 375–390, 2009.

[34]   CIBC, "How to: Send money with Interac e-Transfer - YouTube," 03-Jan-2017. [Online]. Available: https://www.youtube.com/watch?v=X8eIH_KDiz4. [Accessed: 05-Jul-2019].

[35]   E. Johnson, "RBC customer out of pocket after fraud: What you need to know if you e-transfer money | CBC News," *CBC News*, 23-May-2019. [Online]. Available: https://www.cbc.ca/news/business/rbc-customer-out-of-pocket-after-e-transfer-fraud-1.5128114. [Accessed: 23-May-2019].

[36]   J. Jürjens, *Secure systems development with UML*. Dep. of Informatics Software and Systems Engineering, Technische Universität München, Boltzmannstr. 3, 85748 München/Garching, Germany, 2005.

[37]   Interac Association/Acxsys Corporation, "Interac Association and Acxsys Corporation join to form Interac Corp.," *Interac Reports*, 05-Feb-2018. [Online]. Available: https://newsroom.interac.ca/interac-association-and-acxsys-corporation-join-to-form-interac-corp/. [Accessed: 26-Mar-2019].

[38]   Interac Corp., "Our Company," *www.interac.ca*, 2018. [Online]. Available: https://www.interac.ca/en/about/our-company.html. [Accessed: 30-Mar-2019].

[39]   Interac Inc., "Frequently Asked Questions," 2010. [Online]. Available: https://web.archive.org/web/20101205184429/http://www.interac.ca/consumers/faqs.php#e Transfer. [Accessed: 30-Mar-2019].

[40]   Payments Canada, "Spam Emails & Phishing Notice," 2018. [Online]. Available: https://www.payments.ca/about-us/news/spam-emails-phishing-notice. [Accessed: 22-Mar-2019].

## Appendix A

The following is additional information on the history of Interac e-Transfer Companies and Brands:

The Interac e-Transfer platform was originally introduced as *Email Money Transfer* by CertaPay Inc. in June 2002 [3]. In January 2003 CertaPay Inc. was acquired by Acxsys Corporation which was joined with the former not-for-profit Interac Association between 2013 and 2018 to build the for-profit Interac Corp. [37], [38]. Additionally, in 2010 the name of the service was changed to Interac e-Transfer in order to incorporate new mobile banking options as e.g. sending money to mobile phone numbers [39].

## Appendix B

Below follows a detailed description of different types of Interac e-Transfer which are relevant for this paper. Currently three types of Interac e-Transfers can be performed by consumers [5], whereby the original Interac e-Transfer (*Standard e-Transfer*) is performed in the following eight steps:

1. Login:
   The customer (sender) logs in to his financial institutions online or mobile banking and selects the option for sending an Interac e-Transfer
2. Specification of transfer details:
   The sender specifies the amount to be sent, selects the account from which it shall be deducted and also the contact to whom it shall be sent. In the case that this contact is not yet within the contact list, the sender specifies the recipient's details, being the name and e-mail address or mobile phone number. Optionally it is possible to specify a custom message to the recipient.
3. Specification of a security question:
   The sender specifies a security question and answer which the recipient must correctly enter in order to being able to deposit the transferred amount.
4. The sender reviews and sends the e-Transfer.
5. The recipient is notified by an email or SMS from Interac which contains a specific link to choose the financial institute where the amount shall be deposited.
6. After logging in, the recipient is asked to enter the security answer.
7. After correctly entering the security answer, the recipient chooses the account into which the amount shall be deposited. Alternatively, the recipient may also choose to reject the transfer.
8. The sender is notified about the deposited or rejected transfer by e-mail.

An *Interac e-Transfer Autodeposit* (*Autodeposit e-Transfer*) in general is similar to the original Interac e-Transfer but steps 2, 7 (becoming step 4) and 8 (becoming step 5) are different and steps 3, 6 and 7 are omitted:

1. Login:
   The customer (sender) logs in to his financial institutions online or mobile banking and selects the option for sending an Interac e-Transfer.
2. Specification of transfer details:
   The sender specifies the amount to be sent, selects the account from which it shall be deducted and also the contact to whom it shall be sent. In the case that this contact is not yet within the contact list, the sender specifies the recipient's details, being the name and the email address. The email address must be registered for Autodeposit by the recipient in advance. The sender is notified about the activated Autodeposit feature and presented with the legal name of the recipient for verification.
3. The sender reviews and sends the e-Transfer.
4. The amount is automatically deposited into the recipient's account (which was registered for the specified email address). The recipient has no option to reject the transfer.
5. The sender and recipient are notified about the deposited transfer by e-mail.

An *Interac e-Transfer Request Money* (*Request Money e-Transfer*) can be seen to request the initiation of a one-time ad-hoc Autodeposit and consists of the following steps:

1. Login:
   The customer (requestor) logs into his financial institution's online or mobile banking and selects the option for Interac e-Transfer Money Request.
2. Specification of transfer details:
   The requestor specifies the amount to be requested, selects the account to which it shall be deposited and the contact to whom the request shall be sent. In the case that this contact is not yet within the contact list, the requestor specifies the recipient's details, being the name and e-mail address or mobile phone number. Optionally it is possible to specify a custom message to the recipient.
3. The requestor reviews and sends the e-Transfer Money Request.
4. The request recipient is notified by an email or SMS from Interac which contains a specific link to choose the financial institute from where the amount shall be transferred. Alternatively, the recipient may choose to decline the request.
5. After logging in, the request recipient is asked to accept (or to decline) the transfer and to specify from which account the amount shall be transferred.
6. After confirming the details, the amount is automatically transferred into the requestor's account (a security question is not needed).
7. The requestor is notified about the accepted (and deposited) or declined transfer by e-mail.

Appendix C

The following constraints and limitations defined the boundaries of our research:

- The given timeframe of about 2 months from project approval to submission deadline did not allow for large scale experiments which would have included the support of other people in order to send and receive transfers from different accounts at different financial institutions. This would have required an approval of the University of Ottawa's *Research Ethics Board (REB)*, due to possible risks to the participants (e.g. freezing of funds/accounts) and the processing of personal information (including financial information). Since this approval was unlikely to be granted within 2 months, the experiments are limited to 3 accounts held by the authors at three different financial institutes.
- The research was also limited by daily transaction limits which apply to Interac e-Transfers and especially to some student accounts. Another limitation was the availability of liquid funds within the accounts. The experiment design had to take these limitations into account so that e.g. one account only transferred a maximum of CAD 1,000 per day, since this was the daily sending limit on the account.
- At last, the research was limited by the availability of access to mobile phone numbers. Sending Interac e-Transfer notifications to virtual mobile phone numbers, which e.g. are provided by the Android App TextMe, did not succeed (messages were not delivered). Therefore, the experiments were designed in a way that only three different mobile phone numbers, to which the authors had personal access, were used.

## Appendix D

The following screenshots of the web page CheckTLS.com (https://www.checktls.com/TestReceiver) show that emails sent to email-addresses ending in *@bell.net*, *@sympatico.ca*, *@cogeco.ca* and *@videotron.ca* are probably not TLS encrypted in transit (more detailed data, including xml-representations of the CheckTLS results can be found in the experimental dataset within the folder "Experimental Data\Support Data"). The webpage https://www.checktls.com/TestReceiver provides functionality to check for the availability of TLS connections when sending mails to available mail servers at specified domains by establishing a connection to these servers. Therefore, it does not test the availability of TLS when emails are sent from these servers, but this is not relevant for this paper.

**CheckTLS Confidence Factor for "bell.net": 0**

| MX Server | Pref | Answer | Connect | HELO | TLS | Cert | Secure | From |
|---|---|---|---|---|---|---|---|---|
| mxmta.owm.bell.net [184.150.200.210:25] | 0 | OK (18ms) | OK (36ms) | OK (18ms) | FAIL | FAIL | FAIL | OK (90ms) |
| Average | | 100% | 100% | 100% | 0% | 0% | 0% | 100% |

*Figure 5: CheckTLS.com confidence factor for "bell.net" (as receiver)*

**CheckTLS Confidence Factor for "sympatico.ca": 0**

| MX Server | Pref | Answer | Connect | HELO | TLS | Cert | Secure | From |
|---|---|---|---|---|---|---|---|---|
| mxmta.owm.bell.net [184.150.200.82:25] | 0 | OK (23ms) | OK (25ms) | OK (22ms) | FAIL | FAIL | FAIL | OK (92ms) |
| Average | | 100% | 100% | 100% | 0% | 0% | 0% | 100% |

*Figure 6: CheckTLS.com confidence factor for "sympatico.ca" (as receiver)*

**CheckTLS Confidence Factor for "cogeco.ca": 0**

| MX Server | Pref | Answer | Connect | HELO | TLS | Cert | Secure | From |
|---|---|---|---|---|---|---|---|---|
| MX.cogeco.ca [216.221.81.26:25] | 10 | OK (13ms) | OK (748ms) | OK (14ms) | FAIL | FAIL | FAIL | OK (813ms) |
| Average | | 100% | 100% | 100% | 0% | 0% | 0% | 100% |

*Figure 7: CheckTLS.com confidence factor for "cogeco.ca" (as receiver)*

**CheckTLS Confidence Factor for "videotron.ca": 0**

| MX Server | Pref | Answer | Connect | HELO | TLS | Cert | Secure | From |
|---|---|---|---|---|---|---|---|---|
| mx.videotron.ca [24.201.245.37:25] | 10 | OK (14ms) | OK (1,105ms) | OK (15ms) | FAIL | FAIL | FAIL | OK (1,306ms) |
| Average | | 100% | 100% | 100% | 0% | 0% | 0% | 100% |

*Figure 8: CheckTLS.com confidence factor for "videotron.ca" (as receiver)*

## Appendix E

The following screenshot and email header information show (compared to emails sent to address at domain @outlook.com) that emails from Interac sent to email-addresses ending in *@boun.cr* are not TLS encrypted in transit (more detailed data, including an xml-representation of the CheckTLS result can be found in the experimental dataset within the folder "Experimental Data\Support Data"). The https://www.checktls.com/TestReceiver provides functionality to check for the availability of TLS connections when sending mails to available mail servers at specified domains via establishing a connection to these servers. Therefore, it does not test the availability of TLS when emails are sent from these servers, but this is not relevant for this paper:

### CheckTLS Confidence Factor for "boun.cr": 0

| MX Server | Pref | Answer | Connect | HELO | TLS | Cert | Secure | From |
|-----------|------|--------|---------|------|-----|------|--------|------|
| mail.boun.cr [104.131.201.102:25] | 10 | OK (2ms) | OK (125ms) | OK (90ms) | FAIL | FAIL | FAIL | OK (292ms) |
| Average | | 100% | 100% | 100% | 0% | 0% | 0% | 100% |

*Figure 9: CheckTLS.com confidence factor for "boun.cr" (as receiver)*

```
9   Authentication-Results: spf=pass (sender IP is 64.254.22.193)
10  smtp.mailfrom=payments.interac.ca; outlook.com; dkim=pass (signature was
11  verified) header.d=payments.interac.ca;outlook.com; dmarc=bestguesspass
12  action=none header.from=payments.interac.ca;
13  Received-SPF: Pass (protection.outlook.com: domain of payments.interac.ca
14  designates 64.254.22.193 as permitted sender)
15  receiver=protection.outlook.com; client-ip=64.254.22.193;
16  helo=notification.payments.interac.ca;
17  Received: from notification.payments.interac.ca (64.254.22.193) by
18  BY2NAM03FT049.mail.protection.outlook.com (10.152.85.177) with Microsoft SMTP
19  Server (version=TLS1_2, cipher=TLS_ECDHE_RSA_WITH_AES_256_CBC_SHA384) id
20  15.20.1730.9 via Frontend Transport; Fri, 22 Mar 2019 15:07:39 +0000
21  X-IncomingTopHeaderMarker: OriginalChecksum:EB5B046A466D7ED2EF1B59A86101B5DAFCBAAFDB47B458BC10678724
22  Received: from mtlpnot02.prod.certapay.com (unknown [10.5.100.18])
23     by notification.payments.interac.ca (Postfix) with ESMTP id AB5DA48059D
24     for <ebc5389project-receiver-transfer1@outlook.com>; Fri, 22 Mar 2019 11:07:38 -0400 (EDT)
25  DKIM-Signature: v=1; a=rsa-sha256; c=relaxed/simple; d=payments.interac.ca;
26     s=default; t=1553267258;
27     bh=anezNp7hkHJ41xzN7/iqFs+eekVuQZsZ/GRichORxPM=;
28     h=Date:From:Reply-To:To:Subject;
29     b=4wFAMpLIYDusA97noWLI1tJbMm2rKQbUqXYKqQnPyBtjl5Hhu9BEqE0ugCIInEg/a
30     vdmxFtD73ZQaksYZl5v+j+BL/QXsHS/80BGtlZE+I3voVWKiC72q+kFP4Tczui37kX
31     54i6uHVIVdBz7IXwxPjabjzuptLiGe+fQRLbVBzs=
32  Date: Fri, 22 Mar 2019 11:07:38 -0400 (EDT)
33  From: "                    " <notify@payments.interac.ca>
34  Reply-To: "          " <ebc5389project-sender-transfer1@outlook.com>
35  To: "          " <ebc5389project-receiver-transfer1@outlook.com>
36  Message-ID: <416298613.105544396.1553267258700.JavaMail.app_prod@mtlpnot02.prod.certapay.com>
37  Subject: INTERAC e-Transfer:                    sent you money.
```

*Figure 10: Interac e-Transfer notification sent to ebc5389project-receiver-transfer1@outlook.com, **using** TLS*

```
 1  Received: from HE1EUR02HT076.eop-EUR02.prod.protection.outlook.com
 2   (2603:10b6:800::26) by SN6PR08MB4176.namprd08.prod.outlook.com with HTTPS via
 3   SN2PR01CA0058.PROD.EXCHANGELABS.COM; Sun, 24 Mar 2019 02:30:44 +0000
 4  Received: from HE1EUR02FT034.eop-EUR02.prod.protection.outlook.com
 5   (10.152.10.57) by HE1EUR02HT076.eop-EUR02.prod.protection.outlook.com
 6   (10.152.11.226) with Microsoft SMTP Server (version=TLS1_2,
 7   cipher=TLS_ECDHE_RSA_WITH_AES_256_CBC_SHA384) id 15.20.1730.9; Sun, 24 Mar
 8   2019 02:30:43 +0000
 9  Authentication-Results: spf=pass (sender IP is 104.131.201.102)
10   smtp.mailfrom=boun.cr; outlook.com; dkim=fail (signature did not verify)
11   header.d=payments.interac.ca;outlook.com; dmarc=none action=none
12   header.from=payments.interac.ca;
13  Received-SPF: Pass (protection.outlook.com: domain of boun.cr designates
14   104.131.201.102 as permitted sender) receiver=protection.outlook.com;
15   client-ip=104.131.201.102; helo=boun.cr;
16  Received: from boun.cr (104.131.201.102) by
17   HE1EUR02FT034.mail.protection.outlook.com (10.152.10.67) with Microsoft SMTP
18   Server id 15.20.1730.9 via Frontend Transport; Sun, 24 Mar 2019 02:30:43
19   +0000
20  X-IncomingTopHeaderMarker: OriginalChecksum:C3EE1EAAF9800B589FBAEB90C1692A19CF0B44A3EDF9
21  Received: (Haraka outbound); Sat, 23 Mar 2019 22:30:42 -0400
22  Authentication-Results-Original: boun.cr; ;
23  Received: from notification.payments.interac.ca (payments2.interac.ca [64.254.22.193])
24      by boun.cr (Haraka/2.5.0) with ESMTP id BD66FF70-B844-4537-8D52-43D10DDB086F.1
25      envelope-from <notify@payments.interac.ca>;
26      Sat, 23 Mar 2019 22:30:42 -0400
27  Received: from mtlpnot04.prod.certapay.com (unknown [10.5.100.18])
28      by notification.payments.interac.ca (Postfix) with ESMTP id 5AD084801AE
29      for <receiver3@boun.cr>; Sat, 23 Mar 2019 22:30:42 -0400 (EDT)
30  DKIM-Signature: v=1; a=rsa-sha256; c=relaxed/simple; d=payments.interac.ca;
31      s=default; t=1553394642;
32      bh=905HW/ifFslveOGidahko9IxUB90KI/xQrVdpaQaVw8=;
33      h=Date:From:Reply-To:To:Subject:
34      b=sRW4TF8NvOfp3zwDlQzlW5ChAiXxZ6GttuxCyka4g+dF4h9mirRf93JoKv38QA9pt
35       lE1/olY+mpIQssCKtg5qXNOjRdM8uE/xE9TUxZiQrgxPIG/V8X36Awi0pax47cD87T
36       gzkfLdj/vhxIA8B4Pa91Ehm+MIxN0kVjUJUr5BRI=
37  Date: Sat, 23 Mar 2019 22:30:42 -0400 (EDT)
38  From: "▮▮▮▮▮▮▮" <notify@payments.interac.ca>
39  Reply-To: "▮▮▮▮▮▮▮" <sender3@boun.cr>
40  To: "▮▮▮▮▮▮▮" <receiver3@boun.cr>
```

*Figure 11: Interac e-Transfer notification sent to receiver3@boun.cr, **not using** TLS*

## Appendix F

*Table 2: Potentially observable e-Transfer-related information on recipient side (e-Transfer terminator)*

| Type of Transfer | Information sources | Overall Information Visible on Recipient Side without login (Differences highlighted in bold) |
|---|---|---|
| **Standard transfer to email** | Email notification, "Deposit Your Money" - web page | Preferred language, transfer status (sent), **recipient name (custom)**, sender name (sender FI-specific format: custom/legal), amount of transferred money, custom notification message from sender, **(select FI link), expiry date, financial institute of sender**, reference number<br>**Subsequent transfers also contain: preferred FI link (last login for specified email address)** |
| **Standard transfer to mobile phone** | SMS notification, "Deposit Your Money" - web page | Preferred language, transfer status (sent), sender name (likely sender FI-specific format: custom/legal), amount of transferred money, custom notification message from sender, (select FI link), expiry date, reference number |
| **Autodeposit** | Email notification | Preferred language (changes during test were not adapted), transfer status (autodeposit), **recipient name (legal)**, sender name (sender FI-specific format: custom/legal), amount of transferred money, custom notification message from sender, **financial institute of sender**, reference number, **financial institute of recipient** |
| **Money Request to email** | Email notification, "Request For Money" - web page, email confirmation | Preferred language, transfer status (request & deposited), **recipient name (custom and - depending on recipient FI - also legal)**, sender name (sender FI-specific format: **legal/custom & legal**), amount of transferred money, custom notification message from sender, **(select FI link), expiry date, financial institute of sender, sender email, financial institute of recipient,** reference number<br>**Subsequent requests also contain: preferred FI link (last login for specified email address)** |
| **Money Request to mobile phone** | SMS notification, "Request For Money" - web page, email confirmation | Preferred language, transfer status (request & deposited), **recipient name (custom or likely - depending on recipient FI - legal)**, sender name (sender FI-specific format: **legal/custom & legal**), amount of transferred money, custom notification message from sender, **(select FI link), expiry date,** reference number, **sender email, financial institute of recipient** |

# Appendix G

The following four tables show the information within each e-Transfer performed for the privacy experiment. Originally, the Table is one piece, but it is split up to show it here. The relevant descriptive columns are repeated for each part. The test plan and protocol for the experiment are contained in the experimental dataset (within the folder "Experimental Data").

*Table 3: Potentially observable information in e-Transfer notifications (privacy experiment) 4/4*

For the "Information in Confirmation to Recipient (email)" columns, non-Request rows contain the note: *"This type of confirmation is only generated for Interac e-Transfer Requests."*

| Step | Transfer | Repetition | Type of Transfer | Sender Account | Deposit at/Transfer From | Amount (small = CAD 15, medium = CAD 235 to CAD 250, big = CAD 485, year 2018 average is CAD 371) | Sender E-Mail TLS | Recipient E-Mail TLS | time between fund transfer authorization by recipient to deposit confirmation | Display language | Action/status (sent/autodeposited/request) | Recipient Name | Sender Name | Amount | Custom notification message | Recipient FI | Sender email |
|---|---|---|---|---|---|---|---|---|---|---|---|---|---|---|---|---|---|
| 1 | 1 | 1 | Standard transfer to email | Scotiabank | RBC | small | true | true | N/A | | | | | | | | |
| 2 | 1 | 2 | Standard transfer to email | Scotiabank | RBC | small | true | true | N/A | | | | | | | | |
| 3 | 2 | 1 | Standard transfer to email | Scotiabank | RBC | medium | true | true | N/A | | | | | | | | |
| 4 | 2 | 2 | Standard transfer to email | Scotiabank | RBC | medium | true | true | N/A | | | | | | | | |
| 5 | 3 | 1 | Standard transfer to email | Scotiabank | RBC | small | false | false | N/A | | | | | | | | |
| 6 | 3 | 2 | Standard transfer to email | Scotiabank | RBC | big | false | false | N/A | | | | | | | | |
| 7 | 4 | 1 | Standard transfer to email | RBC | BMO | small | true | true | N/A | | | | | | | | |
| 8 | 4 | 2 | Standard transfer to email | RBC | BMO | medium | true | true | N/A | | | | | | | | |
| 9 | 5 | 1 | Standard transfer to mobile phone | RBC | Scotiabank | small | N/A | N/A | N/A | | | | | | | | |
| 10 | 5 | 2 | Standard transfer to mobile phone | RBC | Scotiabank | medium | N/A | N/A | N/A | | | | | | | | |
| 11 | 6 | 1 | Standard transfer to email | RBC | BMO | big | N/A | N/A | N/A | | | | | | | | |
| 12 | 6 | 2 | Standard transfer to email | RBC | BMO | small | true | true | N/A | | | | | | | | |
| 13 | 7 | 1 | Autodeposit (email) | Scotiabank | RBC | small | true | true | N/A | | | | | | | | |
| 13 | 7 | 2 | Autodeposit (email) | Scotiabank | RBC | big | true | true | N/A | | | | | | | | |
| 14 | 7 | 3 | Autodeposit (email) | Scotiabank | RBC | big | false | false | N/A | | | | | | | | |
| 15 | 8 | 1 | Request to email | RBC | Scotiabank | medium | false | false | max | English | true | custom | legal | true | true | true | true |
| 16 | 8 | 2 | Request to mobile phone | RBC | Scotiabank | medium | false | N/A | N/A | English | true | custom | legal | true | true | true | true |
| 26 | 18 | 1 | Autodeposit (email) | BMO | Scotiabank | small | true | true | N/A | | | | | | | | |
| 27 | 18 | 2 | Autodeposit (email) | BMO | Scotiabank | big | true | true | N/A | | | | | | | | |
| 28 | 18 | 3 | Autodeposit (email) | BMO | Scotiabank | small | true | true | N/A | | | | | | | | |
| 28 | 19 | 1 | Request to email | Scotiabank | BMO | small | false | false | min | English | true | legal | both | true | true | true | true |
| 29 | 19 | 2 | Request to email | Scotiabank | BMO | big | false | false | max | English | true | legal | both | true | true | true | true |

| Step | Transfer | Repetition | Type of Transfer | Sender Account | Deposit at/Transfer From | Amount (small = CAD 15, medium = CAD 235 to CAD 250, big = CAD 485, year 2018 average is CAD 371) | Sender E-Mail TLS | Recipient E-Mail TLS | time between fund transfer authorization by recipient to deposit confirmation | Information in Confirmation to Sender (email) | | | | | | | | |
|---|---|---|---|---|---|---|---|---|---|---|---|---|---|---|---|---|---|---|
| | | | | | | | | | | Display language | Action/status (accepted transfer/deposited) | Recipient Name | Sender Name | Amount | Custom notification message | Custom confirmation message | Sender FI | Reference number |
| 1 | 1 | 1 | Standard transfer to email | Scotiabank | RBC | small | true | true | N/A | English | true | custom | custom | true | true | true | true | false |
| 2 | 1 | 2 | Standard transfer to email | Scotiabank | RBC | small | true | true | N/A | English | true | custom | custom | true | true | true | true | false |
| 3 | 2 | 1 | Standard transfer to email | Scotiabank | RBC | medium | true | true | N/A | English | true | custom | custom | true | true | true | true | false |
| 4 | 2 | 2 | Standard transfer to email | Scotiabank | RBC | medium | true | true | N/A | English | true | custom | custom | true | true | true | true | false |
| 5 | 3 | 1 | Standard transfer to email | Scotiabank | RBC | small | false | false | N/A | English | true | custom | custom | true | true | true | true | false |
| 6 | 3 | 2 | Standard transfer to email | Scotiabank | RBC | big | false | false | N/A | English | true | custom | custom | true | true | true | true | false |
| 7 | 4 | 1 | Standard transfer to email | RBC | BMO | small | N/A | N/A | N/A | English | true | custom | legal | true | true | false (n.c.) | true | false |
| 8 | 4 | 2 | Standard transfer to email | RBC | BMO | medium | N/A | N/A | N/A | Confirmation was only visible within the bank web page. | | | | | | | | |
| 9 | 5 | 1 | Standard transfer to mobile phone | RBC | Scotiabank | small | N/A | N/A | N/A | | | | | | | | | |
| 10 | 5 | 2 | Standard transfer to mobile phone | RBC | Scotiabank | big | N/A | N/A | N/A | | | | | | | | | |
| 11 | 6 | 1 | Standard transfer to email | RBC | BMO | small | N/A | N/A | N/A | | | | | | | | | |
| 12 | 6 | 2 | Standard transfer to email | RBC | BMO | big | N/A | N/A | N/A | | | | | | | | | |
| 13 | 7 | 1 | Autodeposit (email) | Scotiabank | RBC | small | true | true | max | English | true | legal | custom | true | N/A | N/A | true | true |
| 14 | 7 | 2 | Autodeposit (email) | Scotiabank | RBC | big | true | true | max | English | true | legal | custom | true | N/A | N/A | true | true |
| 15 | 8 | 3 | Autodeposit (email) | Scotiabank | RBC | medium | false | false | max | English | true | custom | custom | true | true | true | true | true |
| 16 | 8 | 1 | Request to email | RBC | Scotiabank | medium | false | N/A | N/A | English | true | custom | legal | true | true | true | true | false |
| 26 | 18 | 1 | Autodeposit (email) | BMO | Scotiabank | small | true | N/A | N/A | English | true | both | legal | true | N/A | N/A | true | true |
| 27 | 18 | 2 | Autodeposit (email) | BMO | Scotiabank | big | true | N/A | N/A | English | true | both | legal | true | N/A | N/A | true | true |
| 28 | 18 | 3 | Autodeposit (email) | BMO | Scotiabank | small | true | N/A | N/A | English | true | both | legal | true | N/A | N/A | true | true |
| 28 | 19 | 1 | Request to email | Scotiabank | BMO | small | false | false | min | English | true | custom | custom | true | true | true | true | false |
| 29 | 19 | 2 | Request to email | Scotiabank | BMO | big | false | false | max | English | true | custom | custom | true | true | true | true | false |

*Table 5: Potentially observable information in e-Transfer notifications (privacy experiment) 2/4*

| Step | Transfer | Repetition | Type of Transfer | Sender Account | Deposit at/Transfer From | Amount (small = CAD 15, medium = CAD 235 to CAD 250, big = CAD 485, year 2018 average is CAD 371) | Sender E-Mail TLS | Recipient E-Mail TLS | Display language | Action/status | Recipient Name | Sender Name | Amount | Custom notification message | Personalized FI Link (last login FI) | Selectable FI Links | Expiry Date | Sender FI | Reference number | Sender email |
|---|---|---|---|---|---|---|---|---|---|---|---|---|---|---|---|---|---|---|---|---|
| 1 | 1 | 1 | Standard transfer to email | Scotiabank | RBC | small | true | true | English | true | false | custom | true | true | false | true | true | false | true | false |
| 2 | 1 | 2 | Standard transfer to email | Scotiabank | RBC | small | true | true | English | true | false | custom | true | true | false | true | true | false | true | false |
| 3 | 2 | 1 | Standard transfer to email | Scotiabank | RBC | medium | true | false | English | true | false | custom | true | true | false | true | true | false | true | false |
| 4 | 2 | 2 | Standard transfer to email | Scotiabank | RBC | medium | true | true | English | true | false | custom | true | true | false | true | true | false | true | false |
| 5 | 3 | 1 | Standard transfer to email | Scotiabank | RBC | small | false | false | English | true | false | custom | true | true | false | true | true | false | true | false |
| 6 | 3 | 3 | Standard transfer to email | Scotiabank | RBC | big | false | false | English | true | false | legal | true | true | false | true | true | false | true | false |
| 7 | 4 | 1 | Standard transfer to email | RBC | BMO | small | N/A | N/A | English | true | false | legal | true | true | false | true | true | false | true | false |
| 8 | 4 | 2 | Standard transfer to email | RBC | BMO | medium | N/A | N/A | English | true | false | legal | true | true | false | true | true | false | true | false |
| 9 | 5 | 1 | Standard transfer to mobile phone | RBC | ScotiaBank | small | N/A | N/A | English | true | false | legal | true | true | false | true | true | false | true | false |
| 10 | 5 | 2 | Standard transfer to mobile phone | RBC | ScotiaBank | medium | N/A | N/A | English | true | false | legal | true | true | false | true | true | false | true | false |
| 11 | 6 | 1 | Standard transfer to email | RBC | BMO | small | N/A | N/A | English | true | false | legal | true | true | false | true | true | false | true | false |
| 12 | 6 | 2 | Standard transfer to email | RBC | BMO | big | N/A | N/A | English | true | false | legal | true | true | false | true | true | false | true | false |
| 13 | 7 | 1 | Autodeposit (email) | Scotiabank | RBC | small | true | true | Autodeposits do not need a deposit/authorize web page | | | | | | | | | | | |
| 13 | 7 | 2 | Autodeposit (email) | Scotiabank | RBC | big | true | true | | | | | | | | | | | | |
| 14 | 7 | 3 | Autodeposit (email) | Scotiabank | RBC | big | false | false | | | | | | | | | | | | |
| 15 | 8 | 1 | Request to email | RBC | ScotiaBank | medium | false | false | English | true | false | legal | true | true | false | true | true | false | true | false |
| 16 | 8 | 2 | Request to mobile phone | RBC | ScotiaBank | medium | N/A | N/A | English | true | false | legal | true | true | false | true | true | false | true | false |
| 26 | 18 | 1 | Autodeposit (email) | BMO | Scotiabank | small | true | true | Autodeposits do not need a deposit/authorize web page | | | | | | | | | | | |
| 27 | 18 | 2 | Autodeposit (email) | BMO | Scotiabank | big | true | true | | | | | | | | | | | | |
| 28 | 18 | 3 | Autodeposit (email) | BMO | Scotiabank | small | true | true | | | | | | | | | | | | |
| 28 | 19 | 1 | Request to email | Scotiabank | BMO | small | false | false | English | true | false | both | true | true | false | true | true | false | true | true |
| 29 | 19 | 2 | Request to email | Scotiabank | BMO | big | false | false | English | true | false | both | true | true | false | true | true | false | true | true |

| Step | Transfer | Repetition | Type of Transfer | Sender Account | Deposit at/Transfer From | Amount (small = CAD 15, medium = CAD 235 to CAD 250, big = CAD 485, year 2018 average is CAD 371) | Sender E-Mail TLS | Recipient E-Mail TLS | time between fund transfer initiation by sender to recipient notification (about available funds/deposited funds) | time between fund transfer authorization by recipient to deposit confirmation (min: <2 minutes, max: close to 30 minutes) | Display language | Action/status (sent/autodeposited/request) | Recipient Name | Sender Name | Sender email | Amount | Custom notification message | Personalized FI Link (last login FI) | Select FI Link | Expiry Date | Sender FI | Reference number | Recipient FI |
|---|---|---|---|---|---|---|---|---|---|---|---|---|---|---|---|---|---|---|---|---|---|---|---|
| 1 | 1 | 1 | Standard transfer to email | Scotiabank | RBC | small | true | true | min | N/A | English | true | custom | custom | false | true | true | true | true | true | true | false | N/A |
| 2 | 1 | 2 | Standard transfer to email | Scotiabank | RBC | small | true | true | min | N/A | English | true | custom | custom | false | true | true | true | true | true | true | false | N/A |
| 3 | 2 | 1 | Standard transfer to email | Scotiabank | RBC | medium | true | true | max | N/A | English | true | custom | custom | false | true | true | true | true | true | true | false | N/A |
| 4 | 2 | 2 | Standard transfer to email | Scotiabank | RBC | medium | true | true | min | N/A | English | true | custom | custom | false | true | true | true | true | true | true | false | N/A |
| 5 | 3 | 1 | Standard transfer to email | Scotiabank | RBC | small | false | false | min | N/A | English | true | custom | custom | false | true | true | true | true | true | true | false | N/A |
| 6 | 3 | 2 | Standard transfer to email | Scotiabank | RBC | big | N/A | N/A | min | N/A | English | true | custom | legal | false | true | true | false | true | true | true | false | N/A |
| 7 | 4 | 1 | Standard transfer to email | RBC | BMO | small | true | true | min | N/A | English | true | custom | legal | false | true | true | false | true | false | true | false | N/A |
| 8 | 4 | 2 | Standard transfer to email | RBC | BMO | medium | N/A | N/A | min | N/A | English | true | false | legal | false | true | true | false | true | true | true | false | N/A |
| 9 | 5 | 1 | Standard transfer to mobile phone | RBC | ScotiaBank | small | N/A | N/A | N/A | N/A | English | true | custom | legal | false | false | false | false | true | N/A | false | false | N/A |
| 10 | 5 | 2 | Standard transfer to mobile phone | RBC | ScotiaBank | small | N/A | N/A | N/A | N/A | English | true | false | legal | false | false | false | N/A | N/A | N/A | true | false | N/A |
| 11 | 6 | 1 | Standard transfer to email | RBC | BMO | small | N/A | N/A | min | N/A | English | true | custom | legal | false | true | true | N/A | N/A | N/A | true | false | N/A |
| 12 | 6 | 2 | Standard transfer to email | RBC | BMO | big | true | true | min | N/A | English | true | legal | legal | false | true | true | N/A | N/A | N/A | true | false | N/A |
| 13 | 7 | 1 | Autodeposit (email) | Scotiabank | RBC | small | true | true | min | max | English | true | legal | custom | true | true | true | true | true | true | true | false | true |
| 14 | 7 | 2 | Autodeposit (email) | Scotiabank | RBC | big | true | true | max | max | English | true | legal | custom | false | true | true | N/A | N/A | N/A | true | false | true |
| 15 | 7 | 3 | Autodeposit (email) | Scotiabank | RBC | big | true | true | max | min | English | true | legal | legal | false | true | true | N/A | N/A | N/A | true | false | true |
| 16 | 8 | 1 | Request to email | RBC | ScotiaBank | medium | false | false | N/A | N/A | English | true | false | legal | false | true | false | false | true | true | false | true | true |
| 26 | 8 | 2 | Request to mobile phone | RBC | ScotiaBank | medium | true | true | N/A | max | English | true | legal | legal | false | true | false | N/A | N/A | N/A | true | true | true |
| 27 | 18 | 1 | Autodeposit (email) | BMO | ScotiaBank | small | true | true | min | N/A | English | true | legal | legal | false | true | true | N/A | N/A | N/A | true | true | true |
| 28 | 18 | 2 | Autodeposit (email) | BMO | ScotiaBank | big | true | true | max | N/A | English | true | legal | legal | true | true | true | N/A | N/A | N/A | true | true | true |
| 28 | 18 | 3 | Autodeposit (email) | BMO | ScotiaBank | small | false | false | min | N/A | English | true | legal | legal | true | true | true | false | true | N/A | true | true | true |
| 29 | 19 | 1 | Request to email | Scotiabank | BMO | small | false | false | N/A | min | English | true | custom | both | true | true | true | false | true | true | true | false | true |
| 29 | 19 | 2 | Request to email | Scotiabank | BMO | big | false | N/A | N/A | max | English | true | custom | both | true | true | true | true | true | true | true | false | N/A |

*Information in Notification to Recipient (email/SMS)*

Appendix H

The following is a screenshot from Payments Canada (2018) showing that e-Transfer-related phishing attacks are already reality:

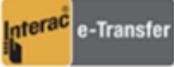

## Spam Emails & Phishing Notice

**Published: June 15, 2018**

New emails using Payments Canada's name or brand are circulating to the Canadian public, inviting users to click a link in order to receive and accept an Interac e-transfer. Please know that **Payments Canada has no connection** to these emails and that they should be deleted right away.

**SAMPLE**

**From:** e-Transfer <id.system47569@payments.ca>
**Sent:** June 11, 2018 12:39 PM
**To:**
**Subject:** Payment Agreement Confirmation

e-Transfer          View in browser   |   Français

**Hi**                                    **,**

Canada Revenue Agency sent you an INTERAC e-Transfer (previously INTERAC Email Money Transfer) $458.00 CAD.

Deposit your money

**Expires:** June 15, 2018

FAQs   |   This is a secure transaction

**INTERAC e-Transfer**
The smart, secure way to send your own money.

© 2000 - 2018 Acxsys Corporation.
All rights reserved. Terms of Use
® Trade-Mark of Interac Inc. Used under licence

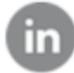 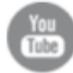

*Figure 12: Spam Emails & Phishing Notice regarding Interac e-Transfers taken from* [40]

Appendix I

The following screenshots show that BMO, RBC, and Scotiabank use security questions to verify the identity of users when they login from a new device or browser.

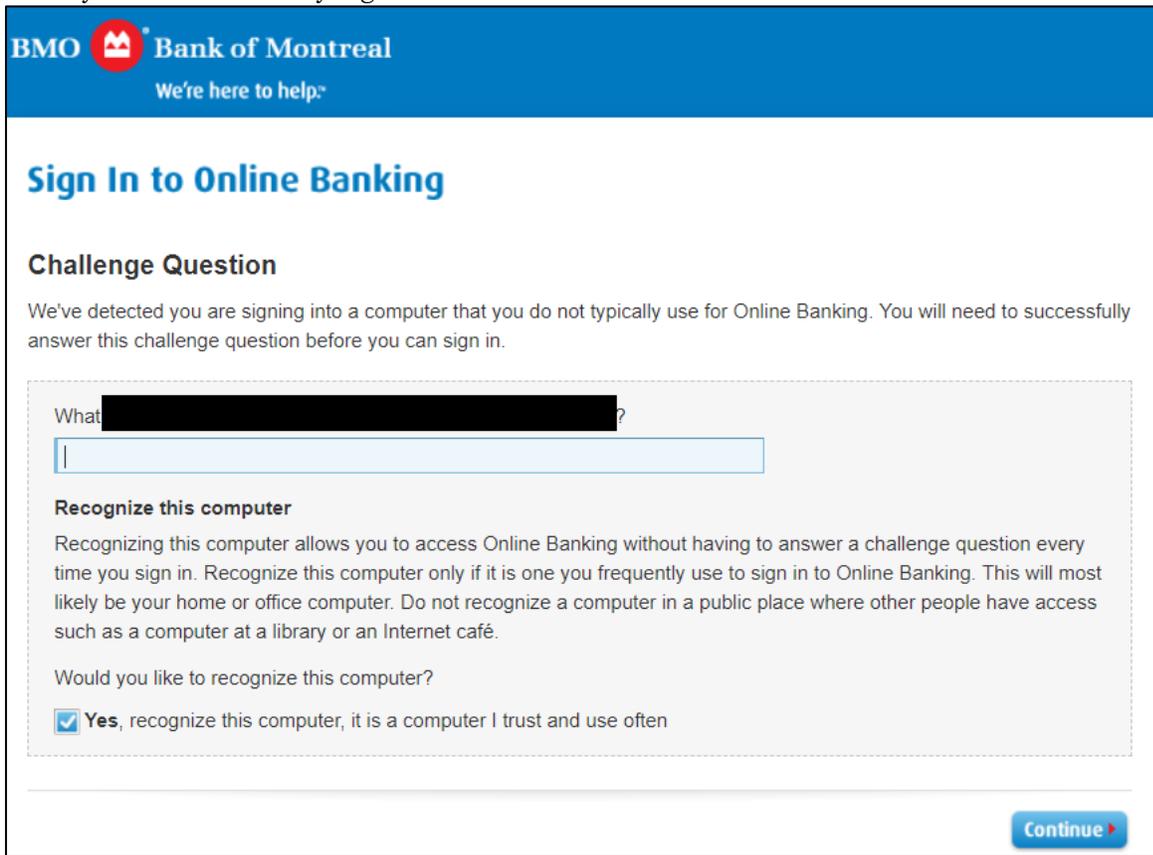

*Figure 13: Screenshot from BMO online banking login when using new device or browser*

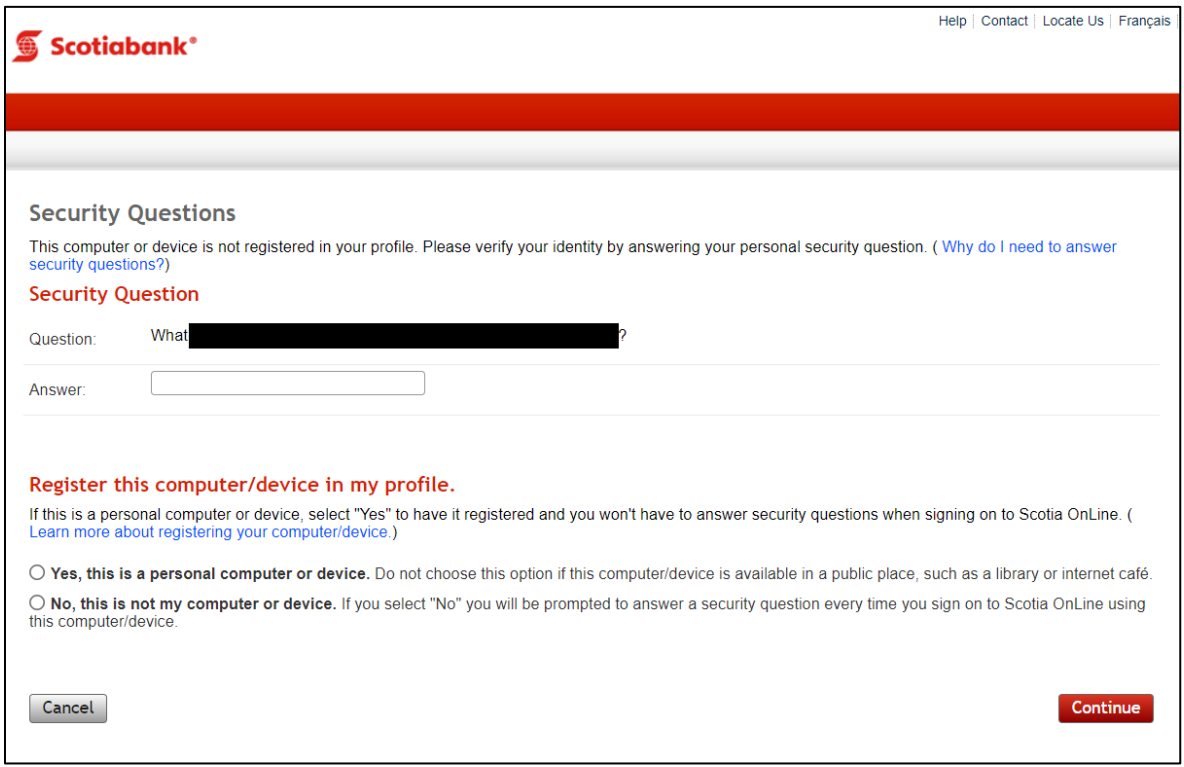

*Figure 14: Screenshot from Scotiabank online banking login when using new device or browser*

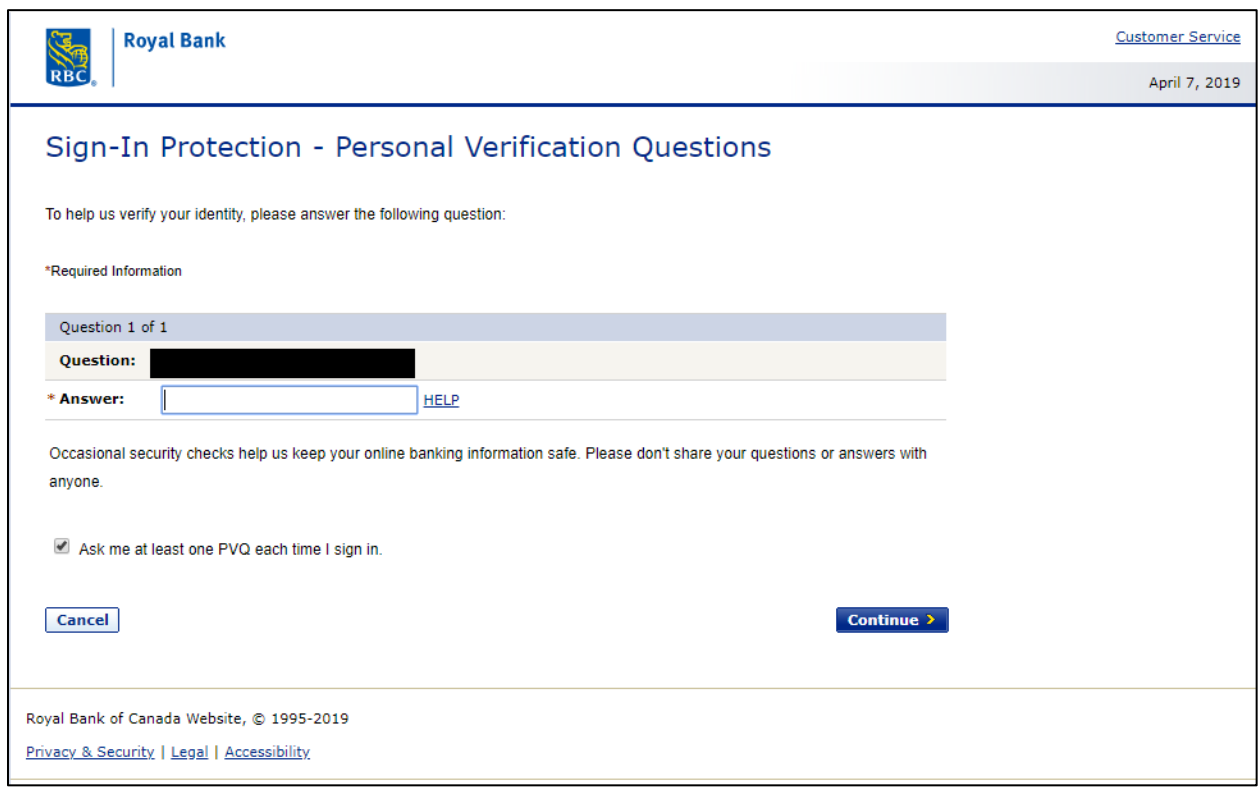

*Figure 15: Screenshot from RBC online banking login when using new device or browser*

# Appendix J

The following three tables show the results for each e-Transfer performed for the security experiment. Originally, the Table is one piece, but it is being split up to show it here. The relevant descriptive columns are repeated for each part.

*Table 7: Results of e-Transfers performed for security experiment 3/3*

| Step | T | Attempted/modified recipient legal name | Recipient (strong/local) IP Address | Confirmation Message | Notes | Time proposed/failed | Time when confirmation / (e) notification received |
|---|---|---|---|---|---|---|---|
| 17 | 9 | <Second Author Legal Name> | <Second author tethering IP> | not configurable | | 2019-03-27 12:11 | 2019-03-27 12:11 |
| 18 | 10 | <Second Author Legal Name> | <Second author tethering IP> | not configurable | | 2019-03-27 12:36 | 2019-03-27 12:36 |
| 19 | 11 | N/A | <First author default IP> | Hi <Second author first name>, you are welcome! <First author first name> | | 2019-03-27 13:00 | 2019-03-27 13:00 |
| 20 | 12 | N/A | <Second author tethering IP> | N/A | Security question gets updated, when sending other transfer, several looks at question, wrong answers count over all transfers, transfer deliberately failed, addresse did not get e-mail regarding cancelation | 2019-03-28 13:47 | 2019-03-28 13:47 |
| 21 | 13 | N/A | <Second author tethering IP> | N/A | Security question gets updated, when sending other transfer, wrong answers count over all transfers, transfer deliberately failed, addresse did not get e-mail regarding cancelation | 2019-03-28 13:47 | 2019-03-28 13:47 |
| 22 | 14 | N/A | <Second author tethering IP> | N/A | Security question gets updated, when sending other transfer, several looks at question, wrong answers count over all transfers, transfer deliberately failed, addresse did not get e-mail regarding cancelation | 2019-03-28 13:47 | 2019-03-28 13:47 |
| 23 | 15 | <Second Author Legal Name> | <Second author tethering IP> | not configurable | | 2019-03-28 00:46 | 2019-03-28 00:46 |
| 24 | 16 | <Second Author Legal Name> | <Second author tethering IP> | not configurable | | 2019-03-28 01:14 | 2019-03-28 01:14 |
| 25 | 17 | <Second Author Legal Name> | <Second author tethering IP> (new IP) | not configurable | Summary of transfers 12-1 to 14-1 | 2019-03-28 15:26 | 2019-03-28 15:26 |

Table 8: Results of e-Transfers performed for security experiment 2/3

| Step | Transfer | Amount | Time between multi transfer interaction (by wizard) minutes (about 2 minutes max, else to a bit) | Message | Security Question | Security/Response/Answer (in order of attempts list licenses) | Time when notification received | Attempts (Attempt/Deposit id) |
|---|---|---|---|---|---|---|---|---|
| 17 | 9 | 0.10 CAD | min | Hi Michel, this is how an e-mail transfer works! <First Author first name> | What is my name? | <First author first name> | 2019-03-27 12:11 BMO | 1 |
| 18 | 10 | 9.90 CAD | min | Hi William, I send you my part of the dinner bill. Liked it! <First Author first name> | What is your name? | William | 2019-03-27 12:34 BMO | 1 |
| 19 | 11 | 11.59 CAD | min | Hi <First Author first name>, thank you for covering my lunch bill yesterday! <Second author first name> | What is the the answer to question 11-1? | Fqa11OK | 2019-03-27 12:52 RBC | 1 |
| 20 | 12 | 200.00 CAD | min | Hi <First Author first name>, that is the amount for the train tickets! <Second author first name> | What is the the answer to question 12-1? | Fqa20000K, Fqa2000K, (Fqa2000K) | 2019-03-27 13:03 BMO | 2 |
| 21 | 13 | 600.00 CAD | max | Hi <First Author first name>, that is the amount for the rental car! <Second author first name> | What is the the answer to question 13-1? | Fqa6000K | 2019-03-27 13:42 BMO | 1 |
| 22 | 14 | 1,100.00 CAD | max | Hi <First Author first name>, That is my part of the hotel bill! <Second author first name> | What is the the answer to question 14-1? | Fqa110000K, (Fqa11000K) | 2019-03-27 14:48 BMO | 1 |
| 23 | 15 | 112.00 CAD | min | Hello Paul, I think I owe you money :-). <First Author first name> | What is my bank? | RBC, RoyalBankOfCanada | 2019-03-28 12:40 BMO | 2 |
| 24 | 16 | 878.00 CAD | min | Dear Mary, This is the first rent. Thank you. <First Author first name> | How much do you get? | 878, CAD878, 878CAD | 2019-03-28 01:08 BMO | 3 |
| 25 | 17 | 1,900.00 CAD | max | Hi <First Author first name>, That is the amount for the train tickets (200), rental car (600) and the hotel bill (1100)! <Second author first name> | What is the combined answer? | Fqa190K, Fqa190K, Fqa19000K | 2019-03-28 15:19 BMO | 3 |

Table 9: Results of e-Transfers performed for security experiment 1/3

| Step | Transfer | Sender Account | Sender Name | Sender E-Mail | Sender E-MailTS | Sender IP | Time sent (Ottawa time) | Recipient Name | Recipient E-Mail/phone Number | Recipient E-MailTS |
|---|---|---|---|---|---|---|---|---|---|---|
| 17 | 9 | RBC | <First author name> | <First author legitimate email> | true | <First author default IP> | 2019-03-27 12:11 | Michel Tremblay | mtremblay@boun.cr | false |
| 18 | 10 | RBC | <First author name> | <First author legitimate email> | true | <First author default IP> | 2019-03-27 12:33 | William Smith | will_smith@boun.cr | false |
| 19 | 11 | Scotiabank | <Second author name> | <Second author legitimate email> | true | <Second author default IP> | 2019-03-27 12:51 | <First author name> | <firstauthorshortname>@boun.cr | false |
| 20 | 12 | Scotiabank | <Second author name> | <Second author legitimate email> | true | <Second author default IP> | 2019-03-27 13:02 | <First author name> | <firstauthorshortname>@boun.cr | false |
| 21 | 13 | Scotiabank | <Second author name> | <Second author legitimate email> | true | <Second author default IP> | 2019-03-27 13:11 | <First author name> | <firstauthorshortname>@boun.cr | false |
| 22 | 14 | Scotiabank | <Second author name> | <Second author legitimate email> | true | <Second author default IP> | 2019-03-27 14:17 | <First author name> | <firstauthorshortname>@boun.cr | false |
| 23 | 15 | RBC | <First author name> | <First author legitimate email> | true | <First author default IP> | 2019-03-28 12:39 | Paul Gagnon | <First author (l) phone number> | N/A |
| 24 | 16 | RBC | <First author name> | <First author legitimate email> | true | <First author default IP> | 2019-03-28 01:07 | Mary Wilson | marywilson@boun.cr | false |
| 25 | 17 | Scotiabank | <Second author name> | <Second author legitimate email> | true | <Second author default IP> | 2019-03-27 14:17 | <First author name> | <firstauthorshortname>@boun.cr | false |

# Appendix K

*Table 10: Additional observations regarding the e-Transfer platform and participating financial institutions*

| Observation | Tentative Consideration |
|---|---|
| The "view in browser"-feature of email notifications uses an incorrectly configured server certificate (see Appendix L). | The certificate uses a wildcard domain (*.interac.ca) which by many browsers is considered to be insecure and therefore should be fixed in order to avoid user confusion. |
| Overall, we noticed that the implementation of Interac e-Transfers and the adoption of features very much differs between the participating financial institutions. This leads to specific problems as described below, but can also be seen as a general problem. | Interac already specifies guidelines for implementing Interac e-Transfer interfaces but these focus on consistent wording and branding [7]. It might be helpful to also provide guidelines on how to implement certain features, e.g. which information should be contained in certain data fields. If these guidelines already exist a mandatory audit should ensure that the guidelines are followed. |
| At BMO and Scotiabank there is no requirement of reentering a security response when sending a Standard Interac e-Transfer. | This makes typos more probable. Although, changing the security answer (or entering it with the typo at the recipient side, too) is possible, requiring to reenter it might be considered to avoid typos beforehand. |
| Only some banks require to acknowledge knowing the recipient of a request or that one's legal name is going to be sent to the recipient of a Request Money e-Transfer. | All customers should be made aware of the implications of sending a certain type of e-Transfer. |
| When changing the sender email address for a Request Money e-Transfer in the RBC interface, this changes the email address for all Interac-notifications. | This implication is not obvious and should be somehow highlighted to the customer. |
| It is possible to send a Money Request e-Transfer which the recipient cannot fulfill (e.g. CAD 5) due to minimum amount limitations configured by the recipient's financial institution (e.g. CAD 10). | Some financial institutions (e.g. BMO, RBC) allow for sending amounts of CAD 0.01. Therefore, consumers might overcome this problem by sending e-Transfers in both directions, whose difference is the desired amount. Seeing this, it seems to be more reasonable to remove these "minimum" amount limitations for all customers. |
| The maximum length of the email address which can be registered for Autodeposits at Scotiabank is smaller than that for recipient email addresses at BMO. | The maximum length of email addresses should be consistently implemented. |
| The preferred-language-of-recipient-setting for Autodeposit e-Transfers sent from BMO is ignored when generating the notification messages. | This can be related to an implementation problem at BMO or Interac and should be fixed. |
| We noticed that RBC processes transfers of bigger amounts which are sent or requested from an RBC account more quickly than BMO and Scotiabank (see Appendix G and Appendix J). | This is a surprising finding, since all financial institutions use the same platform. It is possible that the funds are purposefully withheld in order to profit from interest. This could be researched in future work. |
| The "Request For Money"-webpage contains this sentence: "You can still send money to <sender name (legal/both)> at **<recipient email/mobile phone number>** using INTERAC e-Transfer." | The sentence should actually read: "You can still send money to <sender name (legal/both)> at **<sender email>** using INTERAC e-Transfer." The mobile phone number of the sender is irrelevant, since the request is always sent with an email address as sender. |

## Appendix L

The following screenshots document the use of an incorrectly designed/configured web page and server certificate for the "View-In-Browser" feature of Interac e-Transfer email notifications.

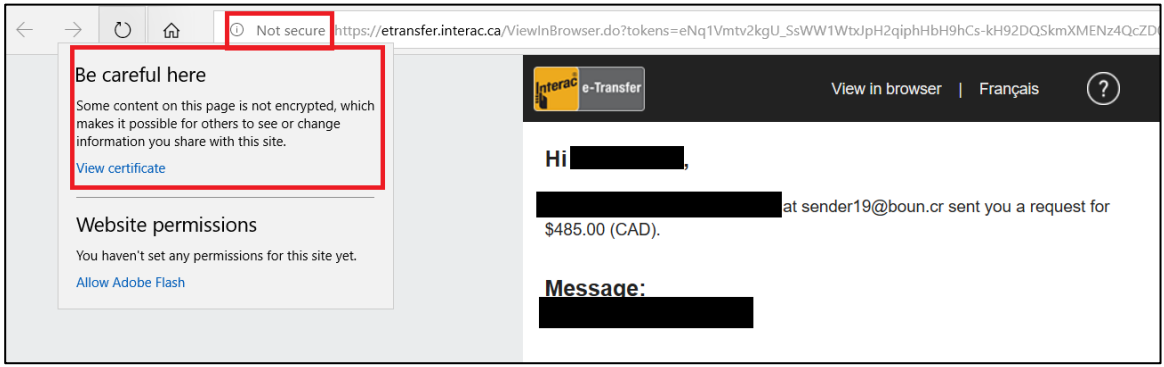

*Figure 16: Warning regarding unencrypted content on "View-In-Browser"-page*

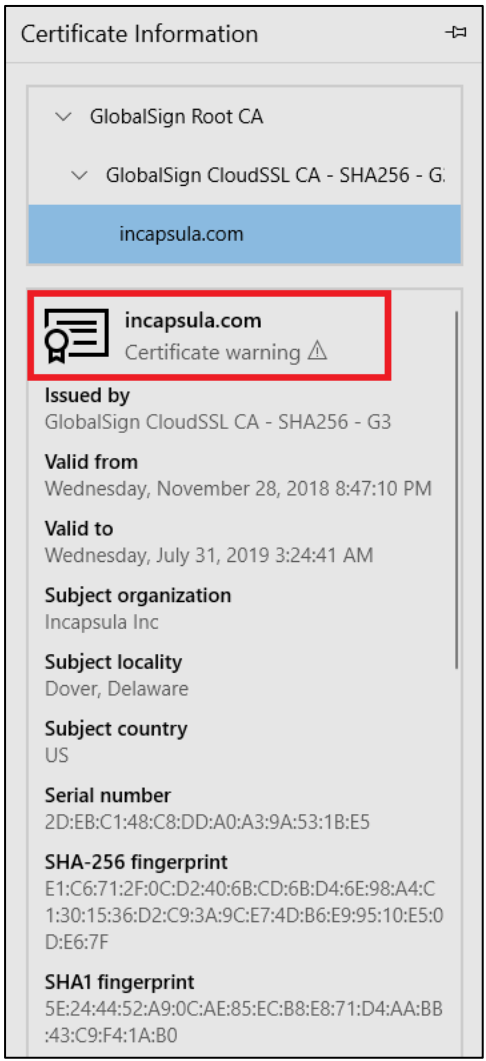

*Figure 18: Certificate warning for "View-In-Browser"-page*

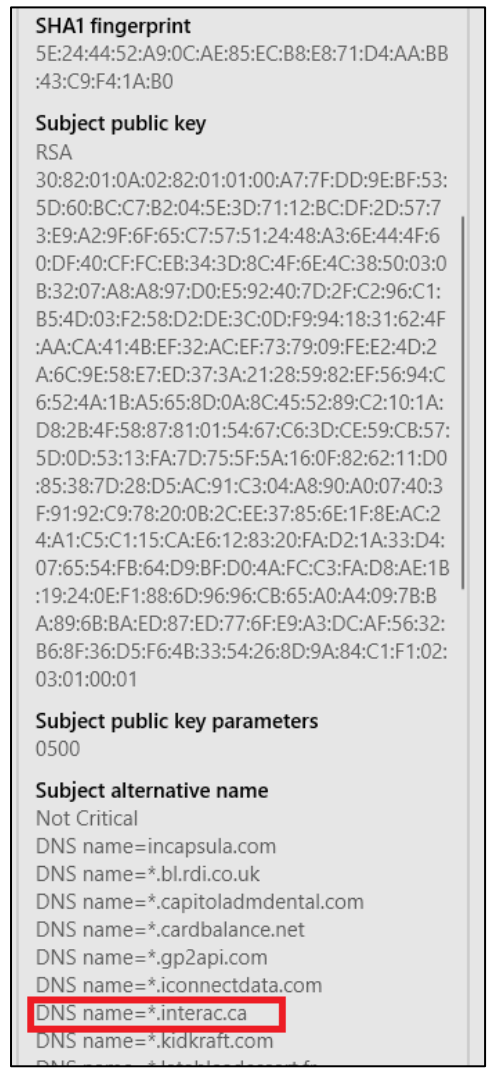

*Figure 17: Wildcard domain certificate for "View-In-Browser"-page*

## Appendix M

The following briefly discusses short term recommendations and proposed improvements to enhance e-Transfer security and privacy (which are related and therefore not grouped), grouped by the parties who would have to implement these. These are intended to overcome or mitigate some of the identified issues until more sophisticated / holistic solutions can be implemented:

### Users:

- Users in general can use Money Request and Autodeposit e-Transfers in order to avoid the use of security questions but they have to be aware that additional private information is then being transmitted.
- To mitigate the risk of eavesdropping (and to address the problem above) users should use an email provider which supports TLS/SSL for incoming and outgoing mail (including the delivery to the user).
- Users should also use virus detectors on their devices to mitigate the risk of being eavesdropped via malware.
- Where it is necessary to use a Standard e-Transfer, users should use *made up information* or real passwords as answers to security questions. The answers should be communicated via a secure channel (as recommend also by Interac Corp.), such as a secure messenger app or in person.
- Users should doublecheck the sender email addresses which they specify for e-Transfers.
- Users should use custom names for Interac e-Transfers (e.g. only first names) wherever they can configure it (not all banks support it for the sender's name).

### Banks and other financial institutions:

- Banks should increase the security of their online banking applications by state-of-the-art two- (or multi-)factor authentication. This would reduce the risk of accounts being hijacked and misused for redirecting e-Transfers.
- Additionally, they can require the authentication of each e-Transfer, e.g. by TAN. This however, does not mitigate the risk of e-Transfers being redirected.
- Banks should monitor the activity of users using the Autodeposit feature in order to identify users who might misuse it for learning email-address/legal name pairs (e.g. based on a high number of subsequent tests of different e-mail addresses).
- Banks should verify the sender addresses of e-Transfers before they can be used (mainly to avoid that users by accident specify wrong addresses).
- Banks can try to support users in finding more secure security questions (e.g. by checking that they do not use obvious questions/answers such as their names). Basic checks can be implemented easily, more sophisticated checks would require more sophisticated natural language processing. The confidentiality of the checked information has to be ensured at all times.
- Banks should more broadly provide the option for users to receive Interac e-Transfer notifications within the online banking portals. Additionally, this could also be integrated into mobile banking applications.
- Banks can try to monitor how many e-Transfers a certain customer looks at in order to identify unusual patterns (resulting from a hijacked account). This however could lead to false positives with negative implications for the respective customer.

**Interac Corp.:**

- Interac might match the addressee names of deposits to the name of the actual person who deposited the amount. However, this can be difficult or impossible to implement due to the use of shortnames and nicknames, different spellings, typos etc.
- If they are capable of doing so, they can also monitor whether a transfer to a certain email address is deposited to an account which belongs to a different person (as compared to the account(s) where previous transfers to the same email address had been deposited). This requires, however, knowledge about the identity of each bank account holder respectively the ability to map different accounts to one person.
- Interac could seek to implement other verification method for direct deposits.
- Interac might mask the information in e-Transfer notifications so that legitimate recipients will nevertheless be able to identify the legitimacy of the notification, but eavesdropper learn less information.
- Alternative (as opt-in for customers) Interac can electronically sign their email notifications so that they can reduce the amount of contained personal information (as this is not required to show the legitimacy of the email, then).
- Interac can create an opt-in selection which allows customers to receive their email notifications in an end-to-end encrypted format.
- Interac (or the banks) might check specified email addresses for TLS support and recommend using another address if it is not supported.
- Interac might end the support for Standard e-Transfers as these are the most insecure. However, the Autodeposit and Money-Request e-Transfers do not overcome all problems as shown in this paper.

**Internet and Email Providers:**

- Some internet providers and email providers have to update/reconfigure their systems in order to support TLS for incoming, outgoing and delivered emails.

# Appendix N

Below diagram maps privacy and security and information problems to proposed additional requirements and improvements to the Interac e-Transfer platform:

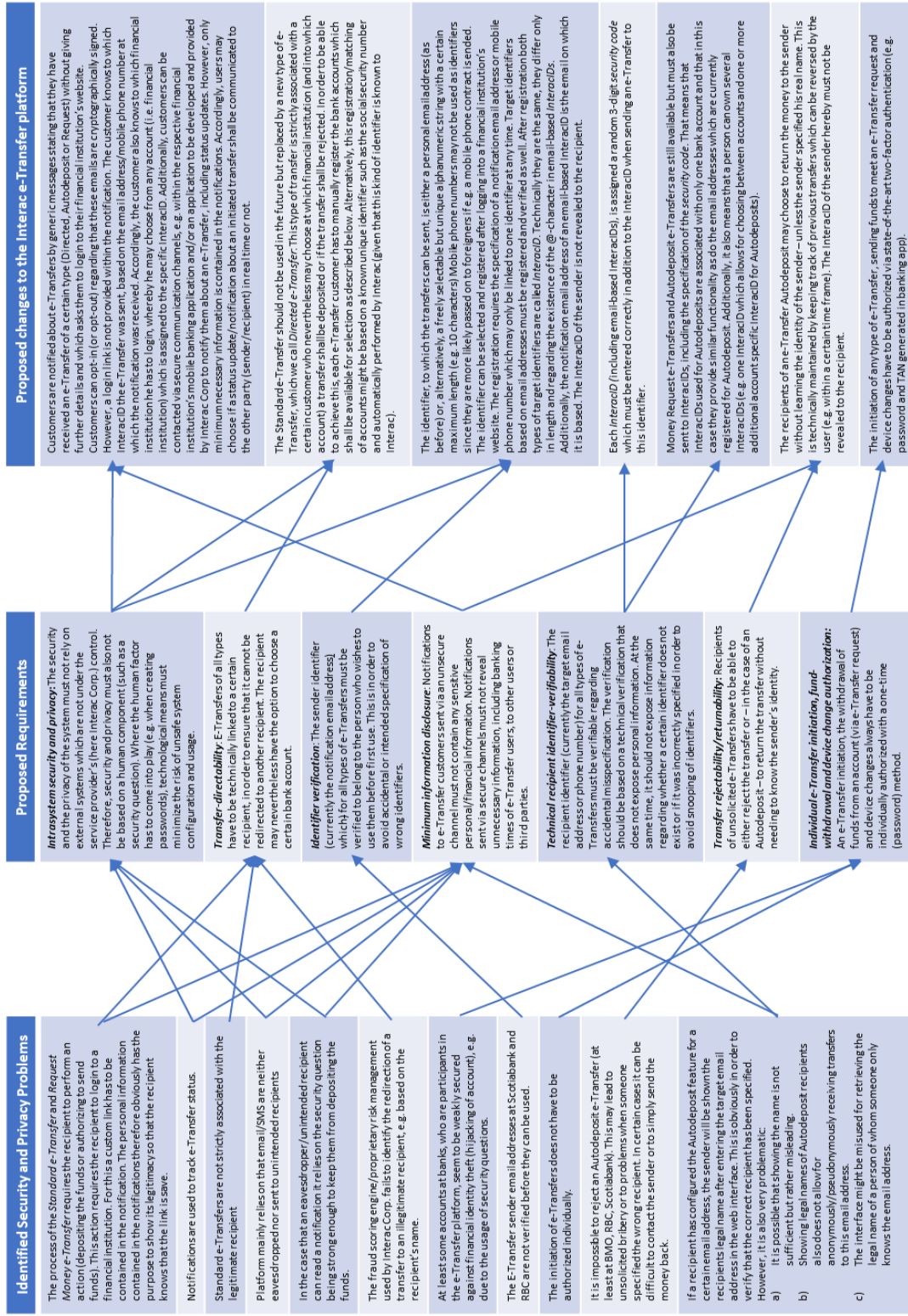

*Figure 19: Mapping of privacy and security problems to proposed requirements and changes to the Interac e-Transfer platform*

Appendix O

The following is the full text (except email signature) from the email we received from Adrienne Vaughan, Senior Manager, External Communications at Interac Corp. on June 3rd, 2019 as reply to our request to review the report/paper before submission:

Hello Fabien and Mohammed,

Thank you for your insightful report about the security and privacy of our *Interac* e-Transfer service. We have shared this study with our subject matter experts and they agree that you have raised some interesting and intelligent points. We are always working towards improving the service that we provide to Canadians – balancing the convenience of frictionless digital payment options with robust privacy and security when transacting online.

As cyber criminal activity rapidly evolves and becomes more sophisticated, we are always evolving our own measures and protocols alongside of it. We also continue to work with all our *Interac* e-Transfer platform participants, law enforcement, including federal and local authorities, regulators, fraud experts and industry stakeholders to protect Canadians, who have placed their trust in us for the secure digital movement of their money.

With this in mind, we would like to point out that we do see some gaps and inaccuracies in your report.

As you can appreciate, our *Interac* e-Transfer platform is our intellectual property. There is proprietary information that is confidential, that we are unable to disclose about how our technology and systems work, the layers of security and fraud protection that we have in place, the advanced technologies that we employ, and how we work with our financial institution partners.

We operate as part of a complex technical ecosystem that includes acquirers, financial institutions and other payment processing entities. The specific details of how we collaborate with them and how responsibilities are allocated between each group is confidential to protect the integrity of the network. Without this operational perspective, which is not available to the general public, it is not possible for external parties to have the full scope of how the *Interac* e-Transfer service works.

Furthermore, you don't have access to the most current information about the service as it's not part of the public domain. For example, the references you use in the report, like the 2002 Canadian Patent, are outdated and do not reflect exactly how the product works today or the security protocols that are in place, as that information is 17 years old.

Similarly, as you point out in the report, there are legal requirements, that you have not considered in your analysis. These requirements, in addition to shareholder and regulatory requirements that we are bound by, are important to how the service is designed and operates, and ultimately result in missing context in your report.

We also should point out that because of how we are connected to Canada's financial institutions, we can see, detect and prevent fraud patterns to keep Canadians protected when sending an *Interac* e-Transfer that are also not apparent to third parties – and as you can appreciate, fraud prevention and protection information does not get into the hands of cybercriminals.

We have made very strategic and deliberate decisions about how the product works and about the privacy, security and fraud protocols that we have put in place. We comply with all applicable privacy laws and regulations.

That said, we agree with you that cyber criminal activity is real and prevalent in Canada and needs to be taken seriously when transacting online for anything, which is why we spend a lot of time and effort working with financial institutions and Canadians to educate them on how they can protect themselves when transacting online. Fighting online crime is a shared responsibility that requires joint efforts of law enforcement, financial institutions and Interac with engagement from the users of the service.

Below are just some of the important tips we share, which also point to some of your report conclusions:
- There are several layers of authentication when a customer moves money online: through online banking authentication, through email authentication, and through *Interac* e-Transfer.
- We continue to stress the importance of using strong and different passwords to access email and online banking services.
- We ask Canadians to choose security questions that are not easy to guess.
- We encourage Canadians to ensure that their email provider uses TLS encryption, or switch to a provider that does.

Finally, we'd like to take the opportunity to correct a few situational facts you've included in your report.
- You referenced a recent CBC article in the appendix of your report, as well as in your email correspondence of May 18, and indicated that we did not answer some of the questions about security and authentication.  That is incorrect.  We provided the statement attached, which shows that we did respond to questions about authentication. However, it was not accurately positioned in the article that way.

- Also, we would like to clarify in your epilogue on page 18, "Due to this and also since we were actually looking into publishing the report, we contacted Interac again and quickly got a response stating that they would like to review the report before publication. Therefore, we incorporated minor changes in the report, e.g. this section and some clarifications based on the feedback we got for the report and the intended submission for publication."  Interac has not yet provided you feedback on your report, as the language used in this sentence may suggest that we already did.

At Interac, we support education studies to improve our products, services and customer experience.  We engage with students on a regular basis.  Generally, they come to us at the beginning with a business problem and we work with them under confidentiality agreements, as our technology and systems are proprietary.

We would appreciate you keeping us updated about the revisions and progress of your report.

Regards,
Adrienne

**Adrienne Vaughan**
**Senior Manager, External Communications [...]**

## Description of Dataset

Our experimental dataset contains the test plan & protocol (including results) and analysis/summary charts in tabular form. Additionally, it contains the raw data information collected during the experiments (emails, pdf-prints of user interfaces, saved web pages). Finally, it also contains screenshots of the online banking user interfaces which were used for preparing this paper, as well as data which proves the statements we made in the paper (such as protocols from CheckTLS.com).

The dataset can be provided in the form of a zip Archive named "Raahemi and Willems - Experimental Dataset.zip" upon request. Its size is 43,495,846 bytes (41 MB). The following hashes have been computed:

SHA1:
F6BF5CECFF829FFF8494DA83D94C031E2F48A422

SHA256:
F97687265169EC37CDC2CCBA7B5BDF65037EF81483DA3DC96769DBCBFFF090CE